\documentclass[lettersize,journal]{IEEEtran}
\usepackage{amsmath,amsfonts}
\usepackage{algorithmic}
\usepackage{algorithm}
\usepackage{array}
\usepackage[caption=false,font=normalsize,labelfont=sf,textfont=sf]{subfig}
\usepackage{textcomp}
\usepackage{stfloats}
\usepackage{url}
\usepackage{verbatim}
\usepackage{graphicx}
\usepackage{cite}
\usepackage{multirow}
\usepackage{microtype}
\begin{document}

\title{A Hybrid Near-field Indoor Channel Model for THz Bands Based on Surface Scattering Characteristics}

\author{Yongchao He,~\IEEEmembership{Member,~IEEE}, Taihao Zhang, Cunhua Pan,~\IEEEmembership{Senior Member,~IEEE}, Hong Ren,~\IEEEmembership{Member,~IEEE}, Chenzhou Lin, Tian Qiu, Bingchang Hua, Cheng-Xiang Wang,~\IEEEmembership{Fellow,~IEEE}, and Jiangzhou Wang,~\IEEEmembership{Fellow,~IEEE}

\thanks{This work was supported by the Key Research and Development Projects under Grant 2023YFB2905100. Y. He, T. Zhang, C. Pan, H. Ren, C. Lin, T. Qiu, and J. Wang are with the National Mobile Communications Research Laboratory, School of Information Science and Engineering, Southeast University, Nanjing 211189, China (e-mail: \{heyongchao, taihao, cpan, hren, 213223687, tianqiu, j.z.wang\}@seu.edu.cn). B. Hua is with Purple Mountain Laboratories, Nanjing, Jiangsu 211111, China (e-mail: huabingchang@pmlabs.com.cn). C.-X. Wang is with the National Mobile Communications Research Laboratory, School of Information Science and Engineering, Southeast University, Nanjing 211189, China; and with the Pervasive Communication Research Center, Purple Mountain Laboratories, Nanjing 211111, China (e-mail: chxwang@seu.edu.cn).}%
} 


\maketitle

\begin{abstract}
Terahertz (THz) communication and extremely large-scale MIMO (XL-MIMO) are essential for achieving ultra-high data rates in future 6G systems. However, at sub-millimeter wavelengths, typical indoor materials exhibit significant roughness that invalidates conventional ideal smooth surface assumptions, while massive array apertures introduce pronounced near-field effects and spatial non-stationarity. To address these challenges, this paper proposes a hybrid near-field channel model utilizing surface scattering characteristics based on distinct measurement campaigns. First, based on typical indoor materials scattering measurements across the 260–400 GHz band, an improved Beckmann-Kirchhoff (B-K) model is developed to accurately characterize surface roughness and diffuse scattering behavior. The model independently analyzes single-bounce (SB) and multi-bounce (MB) clusters by applying deterministic rough surface scattering theory and geometry-statistical approach, respectively. Then, using near-field spatial non-stationarity measurements from a 630-element virtual array in the 330–360 GHz band, a Dual-Gaussian Mixture Model (DMM) and a Negative Binomial (NB) distribution are adopted to describe the lengths and the number of spatial visibility regions (VRs), respectively. Additionally, a Weibull distribution is employed to model the intra-region power fluctuations. Finally, comprehensive XL-MIMO channel evaluations within the same band demonstrate that the proposed model aligns closely with measured results in terms of the spatial cross-correlation function (SCCF), frequency cross-correlation function (FCF), and channel capacity. By reproducing the spatial sparsity of THz band, the proposed model overcomes the limitation of conventional standard models, such as 3GPP 38.901 and WINNER II, in significantly overestimating channel capacity.
\end{abstract}

\begin{IEEEkeywords}
Terahertz, Hybrid Channel Model, Surface Scattering, Near Field, Spatial Non-Stationarity
\end{IEEEkeywords}

\section{Introduction}
\IEEEPARstart{W}{ith} fifth-generation (5G) networks entering global commercial deployment since 2020, research attention has turned to the exploration of sixth-generation (6G) systems \cite{wang2023road, chowdhury20206g, tataria20216g, wang20206g, latvaaho2019key, itu2023imt2030}. Although 5G has achieved significant advancements in enhanced mobile broadband (eMBB) and ultra-reliable and low-latency communications (URLLC), it fails to meet the expected future demands for terabits per second (Tbps) data rates and microsecond-level latency \cite{rappaport2019wireless, Shafie2023terahertz}. Specifically, the capacity of 5G is constrained by its existing sub-6 GHz spectrum resources, restricting its ability to support massive connectivity and ultra-high data traffic \cite{you2021towards}. Therefore, exploring higher frequency bands to acquire vast unallocated spectrum resources is essential. The terahertz (THz) band, generally defined as the spectrum ranging from 0.1 to 10 THz, offers tens of gigahertz of absolute contiguous bandwidth and has emerged as a key technology for 6G and beyond wireless systems research \cite{jiang2024terahertz}. 
THz communication is expected to support various future application scenarios, including Wireless Network-on-Chips (WiNoC) \cite{lee2026multilevel}, the Internet of Nano-Things (IoNT) \cite{chen2024novel}, and Digital Twins \cite{qiu2024terahertz}. Furthermore, the ultra-wide bandwidth and sub-millimeter wavelengths of the THz band facilitate high-capacity data transmission and ultra-high sensing resolution \cite{lyu2026environment}. These unique characteristics make THz technology highly suitable for integrated sensing and communications (ISAC) systems, as well as THz inter-satellite links (THz-ISLs) \cite{dong2026performance, isac2024beam}.

However, the unique propagation mechanisms of THz channels also introduce novel challenges. The extremely high operating frequencies lead to severe free-space path loss and molecular absorption \cite{han2015multi}, while limited diffraction cause THz signals to be vulnerable to blocking obstacles. To compensate for such channel attenuation, deploying extremely large-scale antenna arrays (XL-MIMO) has become a necessary choice. Nevertheless, the combination of massive array apertures and sub-millimeter wavelengths introduces significant near-field effects and spatial non-stationarity (SnS) across the entire array \cite{xu2025spatial}. Additionally, at such short wavelengths, typical indoor materials become electromagnetically rough, which makes the traditional assumption of ideal smooth reflectors inapplicable, causing the coherent specular reflection to attenuate and the incoherent diffuse scattering to become increasingly prominent \cite{taleb2023scattering}. Therefore, existing channel models based on stationarity and ideal smooth surface assumptions are no longer applicable. It is crucial to establish an accurate THz massive near-field scattering channel model for XL-MIMO systems and comprehensively analyze its statistical properties \cite{han2022terahertz}.

Recently, several studies have been conducted in this domain \cite{wang2024far, aljzari2025subthz, wang2024propagation, azpilicueta2023diffuse, serghiou2022terahertz, tarboush2021teramimo, wang2021novel, wang2023novelthz}. \cite{wang2024far} characterized the line-of-sight (LoS) path loss and phase variations across large-scale arrays in the THz band, introducing a cross-field path loss model derived from virtual array measurements. In \cite{aljzari2025subthz}, wideband channel measurements around the 300 GHz band were conducted in diverse indoor and outdoor environments, comprehensively analyzing frequency-dependent channel characteristics such as path loss, delay spread, and coherence bandwidth. Furthermore, \cite{wang2024propagation} measured the monostatic radar cross section (RCS) of typical building materials above 215 GHz, revealing that scattering power heavily depends on surface roughness and incident angles, and severe surface scattering has effectively replaced ideal specular reflection. Building on this, \cite{azpilicueta2023diffuse} systematically measured the dielectric properties of 27 building materials and developed a diffuse-scattering-informed 3D ray-launching (3D-RL) algorithm validated at 300 GHz, confirming rough surface scattering has become a significant characteristic in the THz band.
While the channel characteristics in the THz band have been extensively investigated, translating these findings into analytical channel models remains a critical challenge. Currently, channel modeling methods for the THz band are broadly divided into deterministic and stochastic modeling \cite{serghiou2022terahertz}. Within the stochastic framework, \cite{tarboush2021teramimo} proposed a baseline geometry-based stochastic channel model for THz ultra-massive MIMO system, considering sub-THz molecular absorption and initial sub-array spatial non-stationarity.
To further capture the scattering mechanisms at these frequencies, \cite{wang2021novel} proposed a 3D space-time-frequency non-stationary geometry-based stochastic model (GBSM), introducing the frequency-dependent evolution of diffuse scattering into the intra-cluster parameter generation.
Moreover, \cite{wang2023novelthz} introduced exact spherical wavefronts into the THz channel modeling framework, fundamentally replacing the conventional plane-wave assumption to accurately evaluate near-field spatial characteristics for massive MIMO systems. 

While stochastic models provide preliminary modeling for XL-MIMO systems, their capacity to characterize spatial non-stationarity remains limited, and they fail to explicitly model the specific scattering mechanisms of the THz band \cite{zhou2026general, kim2026power}. Conversely, deterministic models have achieved a series of results in characterizing surface scattering by employing detailed material parameters \cite{ zhang2024deterministic, priebe2011nonspecular, khamse2023scattering, sheikh2020study}. \cite{zhang2024deterministic} used ray tracing to model LoS and specular reflection paths for characterizing THz channel sparsity and near-field propagation, calibrating the simulation parameters with channel measurement data.
Furthermore, \cite{priebe2011nonspecular} implemented the Kirchhoff scattering theory within a ray tracing algorithm to mathematically model non-specular scattering in the THz frequency band.
Similarly, \cite{khamse2023scattering} proposed a modified rough surface scattering model that conserves the energy of the channel impulse response, independent of the scattering tile size.
Besides, \cite{sheikh2020study} included the Beckmann-Kirchhoff (B-K) model in 3D ray tracing simulations to obtain estimates of the impact of surface diffuse scattering on near-field spatial characteristics and channel capacity.
Although deterministic channel models can accurately describe channel characteristics such as surface scattering, they require specific environmental information and involve high computational complexity, making them difficult to be widely applied across diverse scenarios. 

In summary, existing THz channel modeling approaches face significant limitations because stochastic models lack detailed physical scattering mechanisms, while deterministic models suffer from high computational complexity, making it difficult to meet the requirements for future THz communication systems. Additionally, conventional models generally fail to describe the pronounced near-field phase variations induced by massive array apertures.
To bridge this gap, this paper introduces a hybrid near-field channel model based on surface scattering characteristics for THz XL-MIMO systems. 
The model adopts an improved B-K model to describe THz surface scattering behaviors and characterizes the near-field effect through deterministic point-to-point modeling. Moreover, it simplifies the channel geometric representation by separating single-bounce and multi-bounce clusters. Furthermore, to address severe SnS, the evolution of spatial visibility regions (VRs) is mathematically characterized via a Dual-Gaussian Mixture Model (DMM) and a Negative Binomial (NB) distribution. 
The main contributions of this paper are summarized as follows.
\begin{itemize}
    \item An improved B-K scattering model is proposed to accurately describe the electromagnetic properties of typical indoor building materials at sub-millimeter wavelengths. Based on measurement data, this improved model effectively characterizes the differences in power attenuation on both sides of the specular reflection direction, alongside the diffuse background power, thereby overcoming the limitations of traditional smooth surface assumption.
    \item A hybrid near-field indoor channel model based on surface scattering characteristics is established to address the complex propagation mechanisms and near-field effects of THz XL-MIMO systems. This model separates multipath propagation mechanisms by applying deterministic rough surface scattering theory to single-bounce (SB) clusters and a geometry-statistical approach to characterize multi-bounce (MB) clusters. Based on this hybrid framework, the model uses a DMM to describe the lengths of VRs and an NB distribution for the number of VRs. Furthermore, a Weibull distribution is employed to model the intra-region power fluctuations.
    \item The key statistical properties of the proposed channel model, including the spatial cross-correlation function (SCCF), frequency cross-correlation function (FCF), and channel capacity, are theoretically derived and empirically validated. Simulation results in the 330–360 GHz band exhibit a high degree of agreement with measurement data, validating the accuracy of the proposed model. Furthermore, the results indicate that the model accurately describes the spatial sparsity of THz channels, and corrects the overestimation of channel capacity in traditional standard models in sparse scattering environments.
\end{itemize}

The remainder of this paper is organized as follows. Section II introduces the proposed hybrid near-field channel model for THz massive MIMO systems based on surface scattering characteristics. Section III derives the comprehensive statistical properties of the proposed channel model. Section IV details the VNA-based frequency-domain channel measurement system and the experimental setups. Section V provides the detailed measurement results and system performance analysis. Finally, the conclusions are drawn in Section VI.
\section{Channel Model}
\subsection{General Channel Framework}
\begin{figure}[!t]
\centering
\includegraphics[ width=0.8\columnwidth]{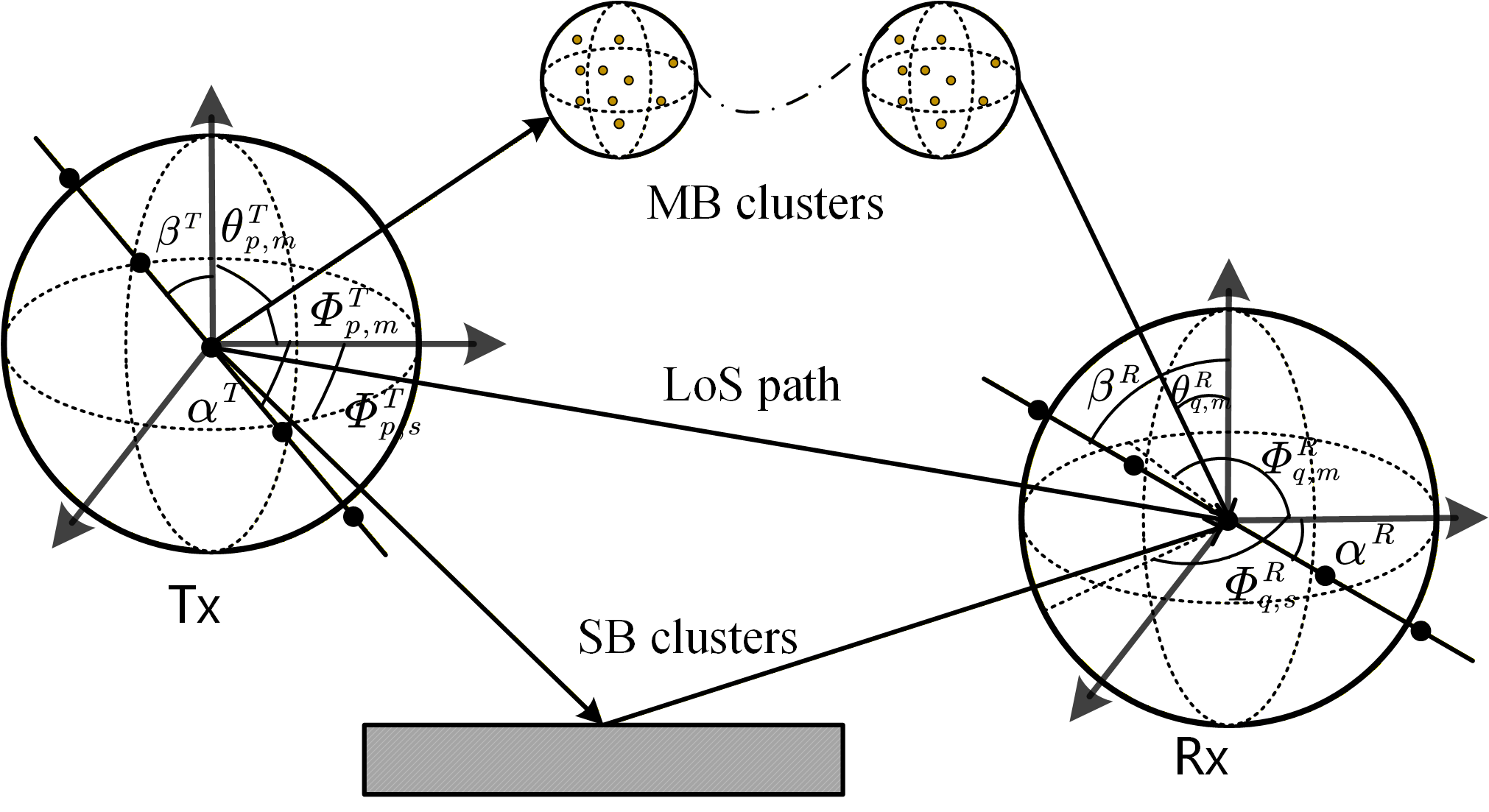}
\caption{Illustration of the proposed hybrid indoor non-stationary near-field channel model for XL-MIMO systems.}
\vspace{-2mm}
\label{model}
\end{figure}

The illustration of the proposed channel model for the XL-MIMO system is presented in Fig. \ref{model}. It is assumed that uniform linear arrays (ULAs) are deployed at both the transmitter (Tx) and receiver (Rx), consisting of $N_T$ and $N_R$ antenna elements with antenna spacings of $d_T$ and $d_R$, respectively. To simplify the model, the effect of the antenna pattern is not considered here, and each array element is assumed to be omnidirectional with unity gain.
The orientations of Tx and Rx are represented by elevation and azimuth angles. The position vectors of the $p$-th transmit antenna $A_p^T$ and the $q$-th receive antenna $A_q^R$ are denoted by $\mathbf{d}_p$ and $\mathbf{d}_q$, respectively. Specifically, assuming the center of each ULA is located at the origin of its local coordinate system, the position vector $\mathbf{d}_p $ of the $p$-th transmit antenna can be expressed as
\begin{equation}
    \mathbf{d}_p = d_p 
    \begin{bmatrix} 
        \cos\beta^T \cos\alpha^T \\ 
        \cos\beta^T \sin\alpha^T \\ 
        \sin\beta^T 
    \end{bmatrix},
\end{equation}
where $\beta^T$ and $\alpha^T$ denote the elevation and azimuth angles of the Tx array, respectively. Similarly, the position vector $\mathbf{d}_q$ of the $q$-th receive antenna is given by
\begin{equation}
    \mathbf{d}_q = d_q 
    \begin{bmatrix} 
        \cos\beta^R \cos\alpha^R \\ 
        \cos\beta^R \sin\alpha^R \\ 
        \sin\beta^R 
    \end{bmatrix},
\end{equation}
where $\beta^R$ and $\alpha^R$ represent the elevation and azimuth angles of the Rx array. Furthermore, to maintain the symmetrical array structure, the scalar distances $d_p$ and $d_q$ from the respective array centers are calculated as
\begin{equation}
    d_p = \frac{N_T - 2p + 1}{2} d_T,\quad d_q = \frac{N_R - 2q + 1}{2} d_R.
\end{equation}
 For subsequent distance calculations, it is necessary to transform these local position vectors into the global coordinate system. Let $\mathbf{r}_{\mathrm{Tx}}$ and $\mathbf{r}_{\mathrm{Rx}}$ denote the absolute 3D spatial coordinates of the geometric centers for the Tx and Rx arrays in the global system, respectively. The absolute spatial position vectors of the $p$-th Tx antenna and the $q$-th Rx antenna, denoted by $\mathbf{r}_{p}^T$ and $\mathbf{r}_{q}^R$, are given by
\begin{equation}
    \mathbf{r}_{p}^T = \mathbf{r}_{\mathrm{Tx}} + \mathbf{d}_p, \quad \mathbf{r}_{q}^R = \mathbf{r}_{\mathrm{Rx}} + \mathbf{d}_q.
\end{equation}

The overall channel frequency response $H_{p,q}$ for each transmit-receive antenna pair $(p,q)$ can be expressed as the product of large-scale path loss and small-scale fading
\begin{equation}
    H_{p,q} = \sqrt{10^{-PL/10}} \cdot \mathrm{FFT}[h_{p,q}],
\end{equation}
where $PL$ represents the large-scale path loss parameter of the channel in dB, given by
\begin{equation}
    PL = 10 \cdot PLE \cdot \log_{10}\left(\frac{d}{d_0}\right) + 20 \log_{10}\left(\frac{4\pi d_0 f}{c_0}\right) + X_{\sigma}.
\end{equation}
Here, $d$ is the distance between the transmitter and receiver, $d_0$ is the reference distance, and $c_0$ is the speed of light. $X_{\sigma}$ denotes the shadow fading that follows the zero-mean Gaussian distribution with deviation of $\sigma_{SF}$. The small-scale fading channel impulse response (CIR) $h_{p,q}(\tau,f_i)$ is modeled as the combination of the LoS, SB, and MB components:
\begin{equation}
\begin{split}
    h_{p,q}(\tau,f_i) &= h_{p,q}^{\mathrm{LoS}}(\tau,f_i) + \sum_{s=1}^{N_{SB}} h_{p,q,s}^{\mathrm{SB}}(\tau,f_i) \\
    &\quad + \sum_{m=1}^{N_{MB}} h_{p,q,m}^{\mathrm{MB}}(\tau,f_i),
\end{split}
\end{equation}
where $N_{SB}$ and $N_{MB}$ denote the numbers of SB and MB clusters, respectively. 
\subsection{Generation of LoS Path Parameters }
The LoS component $h_{p,q}^{\mathrm{LoS}}(\tau,f_i)$ corresponding to the $p$-th Tx antenna and the $q$-th Rx antenna can be expressed as
\begin{equation}
    h_{p,q}^{\mathrm{LoS}}(\tau,f_i) = \sqrt{\frac{K_{p,q}}{K_{p,q}+1}} e^{-j 2\pi f_i \tau_{p,q}^{\mathrm{LoS}}} \delta(\tau - \tau_{p,q}^{\mathrm{LoS}}),
\end{equation}
where $K_{p,q}$ is the Rician factor between antennas $A_p^T$ and $A_q^R$, and the delay $\tau_{p,q}^{\mathrm{LoS}}$ is calculated as $D_{p,q}^{\mathrm{LoS}}/c_0$. Here, $D_{p,q}^{\mathrm{LoS}}$ represents the straight-line distance between them, derived as the Euclidean norm of the distance vector $\mathbf{D}_{p,q}^{\mathrm{LoS}}$,
\begin{equation}
    D_{p,q}^{\mathrm{LoS}} = \left\| \mathbf{D}_{p,q}^{\mathrm{LoS}} \right\|_2,
\end{equation}
where the distance vector $\mathbf{D}_{p,q}^{\mathrm{LoS}}$ from $A_p^T$ to $A_q^R$ is constructed by the geometric vector addition
\begin{equation}
    \mathbf{D}_{p,q}^{\mathrm{LoS}} =  \mathbf{r}_{q}^R-\mathbf{r}_{p}^T  = \mathbf{D} + \mathbf{d}_q - \mathbf{d}_p.
\end{equation}
Here, $\mathbf{D}$ represents the distance vector from the center of the Tx array to the center of the Rx array. The vector $\mathbf{D}$ is defined by the absolute scalar distance $D$ and the LoS spatial angles
\begin{equation}
    \mathbf{D} = D 
    \begin{bmatrix} 
        \cos\theta^{\mathrm{LoS}} \cos\phi^{\mathrm{LoS}} \\ 
        \cos\theta^{\mathrm{LoS}} \sin\phi^{\mathrm{LoS}} \\ 
        \sin\theta^{\mathrm{LoS}} 
    \end{bmatrix},
\end{equation}
where $\theta^{\mathrm{LoS}}$ and $\phi^{\mathrm{LoS}}$ denote the elevation and azimuth angles of the Rx with respect to the Tx, respectively.
\subsection{Generation of SB Cluster Parameters}
In the proposed model framework, the SB clusters constitute the primary source of diffuse multipath components,  acting as the dominant contributors to the non-line-of-sight (NLoS) received energy. Since THz wavelengths are comparable to the roughness of typical indoor surfaces, conventional specular reflection models are inadequate. To address this, the proposed model introduces deterministic ray-tracing geometry to construct the spatial paths, and calculates the multipath energy based on rough surface scattering theory.

For the static indoor environment, the CIR of the $s$-th SB cluster between the $p$-th transmit antenna and the $q$-th receive antenna can be expressed as
\begin{equation}
\begin{split}
    h_{p,q,s}^{\mathrm{SB}}(\tau,f_i) &= \sqrt{\frac{1}{K_{p,q}+1}} \sum_{l=1}^{L_s} \sqrt{P_{p,q,s,l}^{\mathrm{SB}}} \\
    &\quad \times e^{-j\left(2\pi f_i \tau_{p,q,s,l} - \Theta_{s,l}\right)} \delta\left(\tau - \tau_{p,q,s,l}\right)
\end{split},
\end{equation}
where $L_s$ is the number of multipath rays within the $s$-th SB cluster, and $\Theta_{s,l}$ is the initial random phase uniformly distributed over $(0,2\pi]$. $P_{p,q,s,l}^{\mathrm{SB}}$ and $\tau_{p,q,s,l}$ denote the received power and the propagation delay of the $l$-th ray, respectively.

\subsubsection{Scattering Surface Discretization and Delay Calculation}
Cluster-level delays and spatial locations are determined via either generalized stochastic or deterministic approaches, depending on environmental data availability.

In generalized stochastic modeling scenarios with unknown environmental layout, the reference propagation delay $\tau_{s}$ of the $s$-th cluster center, is recursively generated based on the LoS path delay between the Tx and Rx array centers
\begin{equation}
    \tau_{s} = 
    \begin{cases}
        D/c_{0} + \Delta\tau_s, & s = 1 \\
        \tau_{s-1} + \Delta\tau_s, & 2 \le s \le N_{\mathrm{SB}}
    \end{cases},
\end{equation}
where $\Delta\tau_s$ is a normally distributed random variable characterizing the inter-cluster time interval between the $(s-1)$-th and $s$-th SB clusters. Subsequently, by defining $w_s^T$ and $w_s^R$ as the distance ratios of the Tx-to-cluster and cluster-to-Rx segments relative to the total path length, subject to $w_s^T + w_s^R = 1$, the corresponding propagation distances are calculated as $d_s^T = w_s^T \tau_{s} c_0$ and $d_s^R = w_s^R \tau_{s} c_0$, respectively. Given the global position vector of the Tx center $\mathbf{r}_{\mathrm{Tx}}$, along with the azimuth angle of departure (AoD) $\phi_{s}$ and the elevation angle of departure (EoD) $\theta_{s}$, the spatial coordinates of the reflection cluster center, denoted by $\mathbf{r}_s$, is determined as
\begin{equation}
    \mathbf{r}_s = \mathbf{r}_{\mathrm{Tx}} + d_s^T 
    \begin{bmatrix}
        \sin\theta_{s} \cos\phi_{s} \\
        \sin\theta_{s} \sin\phi_{s} \\
        \cos\theta_{s}
    \end{bmatrix}.
\end{equation}

Conversely, if the specific 3D environmental layout is known, the exact spatial coordinates of the cluster centers can be directly obtained using geometric relationships. In such deterministic scenarios, the center coordinate of the $s$-th reflection cluster $\mathbf{r}_s$ serves as a known spatial input extracted directly from the environment. Consequently, the corresponding propagation distances from the $p$-th Tx antenna to the cluster center, and from the cluster center to the $q$-th Rx antenna, are calculated via the Euclidean norms of their position vectors
\begin{equation}
    d_s^T = \left\| \mathbf{r}_s - \mathbf{r}_{\mathrm{Tx}} \right\|_2, \quad d_s^R = \left\|  \mathbf{r}_{\mathrm{Rx}}-\mathbf{r}_s \right\|_2.
\end{equation}
Based on these distances, the corresponding propagation delay for the $s$-th SB cluster is calculated as $\tau_{s} = (d_s^T + d_s^R) / c_0$.

\begin{figure}[!t]
\centering
\includegraphics[width=0.8\columnwidth]{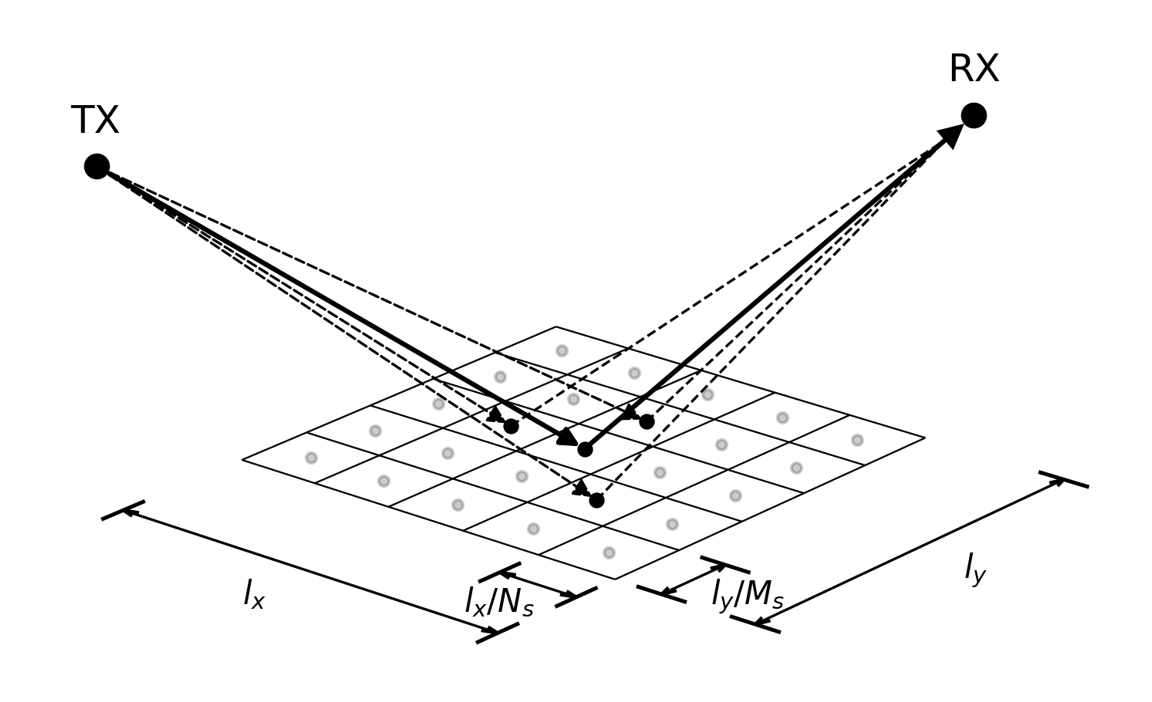}
\caption{Schematic diagram of the surface scattering model.}
\vspace{-2mm}
\label{sca_dia}
\end{figure}
Due to the extremely short wavelength of THz signals, transceivers are often located within the scattering near-field region of rough surfaces. Driven by these prominent near-field effects, the reflecting surface can no longer be treated as an ideal point scatterer. Instead, it is divided into multiple scattering tiles \cite{priebe2011nonspecular}. Accordingly, the total reflecting surface, with dimensions $I_x \times I_y$, is uniformly divided into $N_s \times M_s$ rectangular scattering tiles, as illustrated in Fig. \ref{sca_dia}. Each tile covers an area of $\frac{I_x}{N_s} \times \frac{I_y}{M_s}$ and acts as a secondary scattering source, corresponding to an independent ray within the cluster. To model the intra-cluster multipath rays, a local coordinate system is established with its origin precisely at the acquired cluster center $\mathbf{r}_s$. Let $\mathbf{u}$ and $\mathbf{v}$ denote two orthogonal unit vectors on the local reflecting plane, aligned with the surface's length and width, respectively. The global position vector of the $(i,j)$-th scattering tile, denoted by $\mathbf{r}_{s,i,j}$, is given by
\begin{equation}
    \mathbf{r}_{s,i,j} = \mathbf{r}_s + i \frac{I_x}{N_s} \mathbf{u} + j \frac{I_y}{M_s} \mathbf{v},
\end{equation}
where the indices are bounded by $-\lfloor\frac{N_s-1}{2}\rfloor \le i \le \lceil\frac{N_s-1}{2}\rceil$ and $-\lfloor\frac{M_s-1}{2}\rfloor \le j \le \lceil\frac{M_s-1}{2}\rceil$. 

Once the global position of each tile is established, the propagation delay of the ray traveling from the $p$-th Tx antenna to the $q$-th Rx antenna via the $(i,j)$-th tile is calculated as
\begin{equation}
    \tau_{p,q,s,l} = \frac{\lVert \mathbf{r}_{s,i,j} - \mathbf{r}_p^T \rVert_2 + \lVert \mathbf{r}_{s,i,j} - \mathbf{r}_q^R \rVert_2}{c_0},
\end{equation}
where the ray index $l \in \{1, 2, \dots, L_s\}$ is assigned by sorting all $N_s \times M_s$ calculated delays in ascending order.

\subsubsection{Scattering Coefficients}
\begin{figure}[!t]
\centering
\includegraphics[trim=1mm 1mm 1mm 1mm, clip, width=0.8\columnwidth]{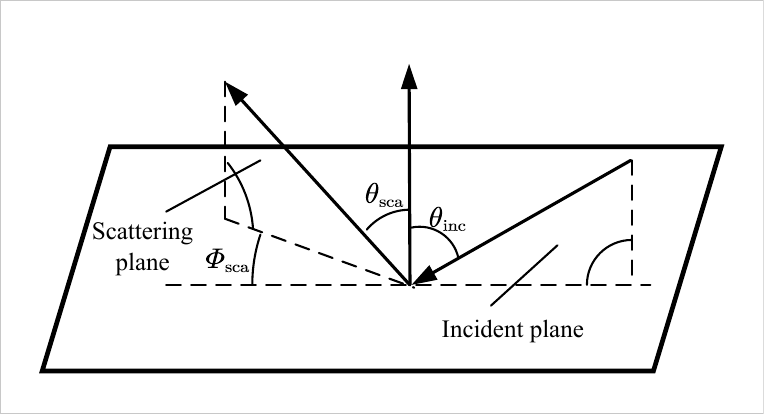}
\caption{Geometry of the three-dimensional surface scattering model.}
\vspace{-2mm}
\label{scatter}
\end{figure}

Based on the local coordinate system, the unit normal vector of the reflecting plane is given by the cross product $\mathbf{n} = \mathbf{u} \times \mathbf{v}$. Then, we define the incident and scattered direction vectors from the scattering tile to the Tx and Rx antennas as $\mathbf{v}_{\mathrm{in}} = \mathbf{r}_p^T - \mathbf{r}_{s,i,j}$ and $\mathbf{v}_{\mathrm{out}} = \mathbf{r}_q^R - \mathbf{r}_{s,i,j}$, respectively. As illustrated in Fig. \ref{scatter}, the elevation incident angle $\theta_{\mathrm{inc}}$, the elevation scattered angle $\theta_{\mathrm{sca}}$, and the relative azimuth scattering angle $\phi_{\mathrm{sca}}$ can be derived as
\begin{align}
    \theta_{\mathrm{inc}} &= \arccos\left( \frac{\mathbf{v}_{\mathrm{in}} \cdot \mathbf{n}}{\left\| \mathbf{v}_{\mathrm{in}} \right\|_2} \right) ,\\
    \theta_{\mathrm{sca}} &= \arccos\left( \frac{\mathbf{v}_{\mathrm{out}} \cdot \mathbf{n}}{\left\| \mathbf{v}_{\mathrm{out}} \right\|_2} \right) ,\\
    \phi_{\mathrm{sca}} &= \arccos\left( \frac{\mathbf{v}_{\mathrm{in}}^{\perp} \cdot \mathbf{v}_{\mathrm{out}}^{\perp}}{\left\| \mathbf{v}_{\mathrm{in}}^{\perp} \right\|_2 \left\| \mathbf{v}_{\mathrm{out}}^{\perp} \right\|_2} \right),
\end{align}
where the geometric projection vectors on the local tangent plane are defined as $\mathbf{v}_{\mathrm{in}}^{\perp} = \mathbf{v}_{\mathrm{in}} - (\mathbf{v}_{\mathrm{in}} \cdot \mathbf{n})\mathbf{n}$ and $\mathbf{v}_{\mathrm{out}}^{\perp} = \mathbf{v}_{\mathrm{out}} - (\mathbf{v}_{\mathrm{out}} \cdot \mathbf{n})\mathbf{n}$.

As each tile acts as a secondary scattering source, its effective area is geometrically constrained by the incident wavefront. Specifically, given the physical area of the individual tile $A = \frac{I_x I_y}{N_s M_s}$, the effective scattering power factor of the $(i,j)$-th scattering tile, denoted by $\eta_{s,i,j}$, is formulated as
\begin{equation}
    \eta_{s,i,j} = |\Gamma|^2 \cdot \langle \rho \rho^* \rangle_{\infty},
\end{equation}
where $\Gamma$ denotes the classical Fresnel reflection coefficient of the surface material. The term $\langle \rho \rho^* \rangle_{\infty}$ represents the average scattering power coefficient \cite{beckmann1987scattering}. According to the improved B-K scattering model, it is calculated as
\begin{equation}
    \langle \rho \rho^* \rangle_{\infty} = e^{-g} \left( \rho_0^2 + \frac{\pi F^2 l_c^2}{ A} \sum_{m=1}^{\infty} \frac{g^m}{m! m} e^{-\frac{(v_x^2 + v_y^2) l_c^2}{4m}} \right) + \Delta,
\end{equation}
where $l_c$ represents the equivalent surface correlation length of the scattering tile under different scattering angles. To reduce modeling complexity, rather than configuring this parameter for each scattering angle, an average equivalent surface correlation length is adopted for each side of the specular reflection angle. Additionally, $\Delta$ accounts for the base diffuse reflection coefficient compensating for background scattering. The roughness factor $g$, describing the electromagnetic phase variance induced by the surface height fluctuations, is given by
\begin{equation}
    g = k^2 \sigma_h^2 (\cos\theta_{\mathrm{inc}} + \cos\theta_{\mathrm{sca}})^2,
\end{equation}
with $k$ being the free-space wavenumber and $\sigma_h$ representing the standard deviation of the surface height profile. Furthermore, the geometric factor $F$, along with the spatial wave vector projections $v_x$ and $v_y$ on the local tile plane, are derived as
\begin{align}
    F &= \frac{1 + \cos\theta_{\mathrm{inc}} \cos\theta_{\mathrm{sca}} - \sin\theta_{\mathrm{inc}} \sin\theta_{\mathrm{scat}} \cos\phi_{\mathrm{scat}}}{\cos\theta_{\mathrm{inc}} (\cos\theta_{\mathrm{inc}} + \cos\theta_{\mathrm{scat}})}, \\
    v_x &= k(\sin\theta_{\mathrm{inc}} - \sin\theta_{\mathrm{sca}} \cos\phi_{\mathrm{sca}}), \\
    v_y &= k(-\sin\theta_{\mathrm{sca}} \sin\phi_{\mathrm{sca}}).
\end{align}
Finally, the coherent scattering component $\rho_0$, which characterizes the specular reflection contribution from the tile, is expressed as
\begin{equation}
    \rho_0 = \mathrm{sinc}\left( v_x \frac{I_x}{N_s} \right) \mathrm{sinc}\left( v_y \frac{I_y}{M_s} \right).
\end{equation}
Here, while $\rho_0^2$ captures the specular reflection from the tile, the subsequent infinite series summation precisely quantifies the power distribution of the non-coherent diffuse scattering components over the spatial domain.

Once the effective scattering power factor for each scattering tile is determined, it is directly mapped to the corresponding $l$-th sorted ray, denoted as $\eta_l$. Given a uniform transmit power $P_0$ per antenna element, the unnormalized received power of the $l$-th ray within the $s$-th SB cluster is calculated as
\begin{equation}
    P_{p,q,s,l}^{\prime} = PL_{p,q,l}^{\mathrm{SB}} \cdot \eta_{l} \cdot P_0,
\end{equation}
where $PL_{p,q,l}^{\mathrm{SB}}$ denotes the path loss associated with the total propagation distance $c_0 \tau_{p,q,s,l}$ of the specific ray.
\subsection{Generation of MB Cluster Parameters}
Unlike SB clusters, MB clusters experience successive unobservable scatterings and reflections within the indoor environment. To avoid the prohibitive computational complexity of precise electromagnetic modeling, a geometry-statistical modeling approach is adopted, where the CIR of the $m$-th MB cluster is formulated as
\begin{equation}
\begin{split}
    h_{p,q,m}^{\mathrm{MB}}(\tau,f_i) &= \sqrt{\frac{1}{K_{p,q}+1}} \sum_{l=1}^{L_m} \sqrt{P_{p,q,m,l}^{\mathrm{MB}}} \\
    &\quad \times e^{-j\left(2\pi f_i \tau_{p,q,m,l} - \Theta_{m,l}\right)} \delta\left(\tau - \tau_{p,q,m,l}\right),
\end{split}
\end{equation}
where $L_m$ denotes the number of resolvable multipath rays within the $m$-th MB cluster, and $\Theta_{m,l}$ is the initial random phase uniformly distributed over $(0, 2\pi]$.
Similar to the SB scenarios, the reference propagation delay $\tau_{m}$ of the $m$-th MB cluster is recursively generated based on the LoS path delay
\begin{equation}
    \tau_{m} = 
    \begin{cases}
        D/c_{0} + \Delta\tau_m, & m = 1 \\
        \tau_{m-1} + \Delta\tau_m, & 2 \le m \le N_{\mathrm{MB}}
    \end{cases},
\end{equation}
here $\Delta\tau_m$ is a normally distributed random variable characterizing the inter-cluster delay difference between the $(m-1)$-th and $m$-th MB clusters.

Subsequently, the physical propagation distances of the MB clusters are geometrically modeled by tracing the observable sub-paths linked to the Tx and Rx arrays. Similar to the SB clusters, the geometric positions of the MB clusters are determined by dividing the total path length into multiple segments. Given $w_m^T$ and $w_m^R$ as the distance ratios for the Tx-to-cluster and cluster-to-Rx links, respectively, the coordinate of the first-bounce scatterer at Tx-side, denoted by $\mathbf{r}_m^T$, is calculated relative to the Tx array center $\mathbf{r}_{\mathrm{Tx}}$ as
\begin{equation}
\renewcommand{\arraystretch}{1.3}
\mathbf{r}_m^T = \mathbf{r}_{\mathrm{Tx}} + w_m^T\tau_{m}c_0
\begin{bmatrix}
\sin\theta_{m}^T \cos\phi_{m}^T \\
\sin\theta_{m}^T \sin\phi_{m}^T \\
\cos\theta_{m}^T
\end{bmatrix},
\end{equation}
where $\theta_{m}^T$ and $\phi_{m}^T$ denote the EoD and the AoD defined at the Tx array center for the $m$-th MB cluster, respectively.  Taking the AoD as an illustrative example, the cluster angle $\phi_{m}^T$ follows a wrapped Gaussian distribution \cite{wang2021angle}, expressed as $\phi_{m}^T = \mathrm{std}[\phi_{m}^T] Y_m + \psi_m$. Here, $Y_m \sim \mathcal{N}(0,1)$ is a standard normal variable, $\mathrm{std}[\phi_{m}^T]$ denotes the standard deviation of the global angular spread for the specific environment, and $\psi_m$ corresponds to the specific reference angle of the $m$-th MB cluster. Thus, the propagation distance from the $p$-th Tx antenna to the $m$-th cluster is calculated as 
\begin{equation}
d_{p,m}^T = \left\|\mathbf{r}_m^T - \mathbf{r}_p^T\right\|_2.
\end{equation}

Let $\mathbf{R}_{p,m} = \mathbf{r}_m^T - \mathbf{r}_p^T$ denote the relative coordinate vector. The corresponding AoD $\phi_{p,m}^T$ and EoD $\theta_{p,m}^T$ are formulated as
\begin{equation}
\phi_{p,m}^T = \arctan\left( \frac{y_{p,m}}{x_{p,m}} \right), \quad \theta_{p,m}^T = \arccos\left( \frac{z_{p,m}}{d_{p,m}^T} \right),
\end{equation}
where $x_{p,m}$, $y_{p,m}$, and $z_{p,m}$ are the Cartesian components of $\mathbf{R}_{p,m}$. And the specific departure azimuth angle of the $l$-th ray within this cluster, denoted by $\phi_{p,m,l}^T$, is given by
\begin{equation}
    \phi_{p,m,l}^T = \phi_{p,m}^T + \Delta\phi_{p,m,l}^T,
\end{equation}
where $\Delta\phi_{p,m,l}^T$ represents the intra-cluster relative angular offset of the $l$-th ray, which follows a zero-mean Gaussian distribution.
Thus, the corresponding directional unit vector from the $p$-th Tx antenna to the $m$-th cluster is defined as
\begin{equation}
    \mathbf{u}_{p,m}^T = [\sin\theta_{p,m}^T\cos\phi_{p,m}^T, \sin\theta_{p,m}^T\sin\phi_{p,m}^T, \cos\theta_{p,m}^T]^T.
\end{equation}
Similarly, the specific directional unit vector for the $l$-th ray is given by
\begin{equation}
    \mathbf{u}_{p,m,l}^T = [\sin\theta_{p,m,l}^T\cos\phi_{p,m,l}^T, \sin\theta_{p,m,l}^T\sin\phi_{p,m,l}^T, \cos\theta_{p,m,l}^T]^T.
\end{equation}

In typical indoor scenarios, the intra-cluster angular spread relative to the cluster center is extremely small. Under such minute angular perturbations, the additional propagation distance discrepancies induced by non-orthogonal tilted scattering planes emerge as negligible high-order infinitesimals. To simplify the model, an orthogonal local scattering plane assumption is adopted. Specifically, the effective scattering plane illuminated by the cluster is assumed to be orthogonal to its mean propagation direction. Consequently, the cosine value of the true spatial deviation angle between the cluster path and the $l$-th specific ray can be formulated via the inner product
\begin{equation}
\begin{split}
    \cos(\Delta\psi_{p,m,l}^T) &= \mathbf{u}_{p,m}^T \cdot \mathbf{u}_{p,m,l}^T \\
    &= \sin\theta_{p,m}^T\sin\theta_{p,m,l}^T\cos(\phi_{p,m}^T-\phi_{p,m,l}^T) \\
    &\quad + \cos\theta_{p,m}^T\cos\theta_{p,m,l}^T
\end{split},
\end{equation} 

Thus, the propagation distance of the $l$-th ray from the Tx antenna to the first scattering plane, denoted by $d_{p,m,l}^T$, is derived as
\begin{equation}
\begin{split}
    d_{p,m,l}^T &= \frac{d_{p,m}^T}{\cos(\Delta\psi_{p,m,l}^T)}.
\end{split}
\end{equation}
Similarly, the propagation distance from the last scattering plane to the Rx antenna for this ray is geometrically calculated and denoted by $d_{q,m,l}^R$. Accounting for the internal reflections, the scaling factors satisfy $ w_m^T +  w_m^R < 1$. 
Therefore, allocating the residual proportion to the unobservable internal bouncing paths, the total propagation distance of the $l$-th ray within the $m$-th MB cluster, denoted by $D_{p,q,m,l}$, is expressed as
\begin{equation}
    D_{p,q,m,l} = d_{p,m,l}^T + d_{q,m,l}^R + (1 - w_m^T - w_m^R)\tau_{m}c_0.
\end{equation}
\subsection{Power Allocation and Normalization of NLoS Components}
The unnormalized power of the $l$-th ray within the $m$-th MB cluster, denoted by $P_{p,q,m,l}^{\prime}$, decays exponentially with respect to its propagation delay \cite{priebe2013power}
\begin{equation}
    P_{p,q,m,l}^{\prime} = \exp\left(-\frac{\tau_{p,q,m,l}}{\Gamma_\tau}\right) \cdot 10^{\frac{Z_m}{10}},
\end{equation}
where $\Gamma_\tau$ represents the delay spread constant of the specific scenario, characterizing the inherent dissipation rate of electromagnetic energy over absolute time. The term $Z_m \sim \mathcal{N}(0, \sigma_Z^2)$ is a zero-mean Gaussian random variable capturing the log-normal shadow fading effect.

To guarantee power consistency within the modeled channel, the total NLoS power must be partitioned between the SB and MB components. The power allocation factor, denoted by $\xi$, is introduced to represent the power ratio of the MB components within the total NLoS energy.
Subsequently, independent local power normalizations are performed for both the SB and MB clusters based on their unnormalized calculated powers.
For the SB clusters, to ensure that the sum of powers across all respective rays exactly equals $1-\xi$, the final normalized power for the $l$-th ray within the $s$-th SB cluster is calculated as
\begin{equation}
    P_{p,q,s,l}^{\mathrm{SB}} = (1-\xi) \cdot \frac{P_{p,q,s,l}^{\prime}}{\sum_{s=1}^{N_{\mathrm{SB}}} \sum_{l=1}^{L_s} P_{p,q,s,l}^{\prime}}.
\end{equation}
Similarly, the final normalized power for the $l$-th ray within the $m$-th MB cluster is constrained by $\xi$ as
\begin{equation}
    P_{p,q,m,l}^{\mathrm{MB}} = \xi \cdot \frac{P_{p,q,m,l}^{\prime}}{\sum_{m=1}^{N_{\mathrm{MB}}} \sum_{l=1}^{L_m} P_{p,q,m,l}^{\prime}}.
\end{equation}
\subsection{Spatial Non-Stationarity and Matrix Formulation}
\label{sec:spatial_matrix_formulation}
In XL-MIMO systems, the significant physical aperture introduces pronounced spatial non-stationarity, which is primarily attributed to the distributed environmental scatterers and dynamic blockages, causing the multipath components to be visible only to specific groups of the antenna elements.
Therefore, an element-specific visibility factor, denoted by $V_{p,q}$, is introduced to characterize the survival state of the spatial link between the $p$-th Tx and $q$-th Rx antennas. Determined by the specific distribution of the environmental scatterers, these distinct visibility factors constitute a spatial visibility matrix $\mathbf{V} \in \mathbb{R}^{N_T \times N_R}$. The visible region length of each consecutive block, represented by $L_v$, follows the DMM, whose probability density function (PDF) is given by
\begin{equation}
f(L_{v}) = \omega_1 \mathcal{N}(\mu_1, \sigma_1^2) + \omega_2 \mathcal{N}(\mu_2, \sigma_2^2),
\end{equation}
where $\omega_1$ and $\omega_2$ are the weighting coefficients representing the occurrence probabilities of different scattering environments, satisfying $\omega_1 + \omega_2 = 1$. The parameters $(\mu_1, \sigma_1)$ and $(\mu_2, \sigma_2)$ characterize the mean and standard deviation of the visible lengths associated with localized environmental features and overall structural elements, respectively. Furthermore, let $N_c$ represent the number of consecutive visible blocks, which follows a NB distribution. By aggregating these blocks along the array axis, a binary spatial visibility matrix $\bar{\mathbf{V}} \in \{0, 1\}^{N_T \times N_R}$ is constructed, where the element $\bar{V}_{p,q} = 1$ indicates that the $m$-th cluster is visible to the link between the $p$-th Tx and $q$-th Rx antennas, and $\bar{\mathbf{V}}_{p,q} = 0$ otherwise.
Then we define the SnS parameter $S$ to characterize the power fluctuations of multipath components across the antenna arrays. This parameter is modeled by a Weibull distribution, i.e., $S \sim \mathcal{W}(A, B)$, where $A$ and $B$ denote the scale and shape parameters, respectively. To ensure physical validity, its values are normalized within the range of $(0, 1]$. Thus, the final spatial visibility matrix $\mathbf{V}$ is derived by mapping the amplitude fluctuations onto the binary mask $\bar{\mathbf{V}}$, which is expressed as
\begin{equation}
V_{p,q} = \bar{V}_{p,q} \cdot S_{p,q}, \quad S_{p,q} \in (0, 1],
\end{equation}

The original stationary channel matrix, denoted by $\mathbf{H}_{\mathrm{stat}} \in \mathbb{C}^{N_T \times N_R}$, is constructed by aggregating the previously derived LoS, SB, and MB components, which is given by
\begin{equation}
    \mathbf{H_{stat}} = 
    \begin{bmatrix}
        h_{1,1} & h_{1,2} & \cdots & h_{1,N_R} \\
        h_{2,1} & h_{2,2} & \cdots & h_{2,N_R} \\
        \vdots  & \vdots  & \ddots & \vdots  \\
        h_{N_T,1} & h_{N_T,2} & \cdots & h_{N_T,N_R}
    \end{bmatrix},
\end{equation} 

The final non-stationary spatial channel matrix $\mathbf{H} \in \mathbb{C}^{N_T \times N_R}$ is generated via the Hadamard product of the spatial visibility matrix and the stationary channel matrix
\begin{equation}
    \mathbf{H} = \mathbf{V} \odot \mathbf{H}_{\mathrm{stat}}.
\end{equation}
\section{Statistical Properties of the Channel Model}
\subsection{SCCF}
The SCCF characterizes the spatial similarity and coherence between distinct spatial sub-links. Specifically, the normalized SCCF between the channel coefficients $h_{p,q}$ and $h_{p',q'}$, is defined as
\begin{equation}
    \rho_{S}(p,q; p',q') = \frac{\mathbb{E} \left[ h_{p,q} \cdot h_{p',q'}^* \right]}{\sqrt{\mathbb{E} \left[ |h_{p,q}|^2 \right] \mathbb{E} \left[ |h_{p',q'}|^2 \right]}},
\end{equation}
where $(\cdot)^*$ denotes the complex conjugate operation, and $\mathbb{E}[\cdot]$ denotes the statistical expectation operator. This expectation is evaluated over independent random variables, including the initial phases, spatial angles, and visibility factors. The unnormalized cross-correlation function can be decomposed into the sum of respective components sub-correlations.
\begin{equation}
    \mathbb{E} \left[ h_{p,q} \cdot h_{p',q'}^* \right] = R_{\mathrm{LoS}} + R_{\mathrm{SB}} + R_{\mathrm{MB}}.
\end{equation}

Specifically, the visibility correlation $R_V(p,q; p',q') = \mathbb{E}[V_{p,q} V_{p',q'}]$ represents the spatial non-stationarity. 
Assuming statistical independence between the visibility blockages and the multipath propagation, the unnormalized spatial cross-correlation for the deterministic LoS component is given by
\begin{equation}
\begin{split}
    R_{\mathrm{LoS}} &= R_V(p,q; p',q') \sqrt{\frac{K_{p,q} K_{p',q'}}{(K_{p,q}+1)(K_{p',q'}+1)}} \\
    &\quad \times \sqrt{P_{p,q}^{\mathrm{LoS}} P_{p',q'}^{\mathrm{LoS}}} \exp\left(-j\frac{2\pi f_c}{c_0} \left(d_{p,q}^{\mathrm{LoS}} - d_{p',q'}^{\mathrm{LoS}}\right)\right)
\end{split}.
\end{equation}
For the NLoS components, let $x \in \{\mathrm{SB}, \mathrm{MB}\}$ denote the specific scattering mechanism, with $n \in \{s, m\}$ representing the corresponding cluster index. The statistical expectations are taken over the mutually independent initial random phases $\Theta_{n,l}^{x}$. Since the expectation $\mathbb{E}[\exp(j\Theta_{n,l}^x - j\Theta_{n',l'}^x)]$ equals $1$ only when $n=n'$ and $l=l'$, and $0$ otherwise, the cross-cluster and cross-ray interference terms vanish. Consequently, the cross-correlation for both SB and MB components simplifies to the coherent superposition of the auto-correlated rays
\begin{equation}
\begin{split}
    R_{x} &= R_V(p,q; p',q') \sqrt{\frac{1}{(K_{p,q}+1)(K_{p',q'}+1)}} \\
    &\quad \times \sum_{n=1}^{N_{x}} \sum_{l=1}^{L_n} \sqrt{P_{p,q,n,l}^{x} P_{p',q',n,l}^{x}} \\
    &\quad \times \exp\left(-j\frac{2\pi f_c}{c_0} \left(D_{p,q,n,l}^{x} - D_{p',q',n,l}^{x}\right)\right)
\end{split}.
\end{equation}
\subsection{FCF}
Given the pronounced frequency selectivity in the THz communications, the normalized FCF between the carrier frequency $f$ and $f + \Delta f$ for the $(p,q)$-th spatial sub-link is defined as
\begin{equation}\label{FCF}
    \rho_F(\Delta f; p,q) = \frac{\mathbb{E} \left[ h_{p,q}(f) \cdot h_{p,q}^*(f + \Delta f) \right]}{\sqrt{\mathbb{E} \left[ |h_{p,q}(f)|^2 \right] \mathbb{E} \left[ |h_{p,q}(f + \Delta f)|^2 \right]}}.
\end{equation}

Similar to the spatial correlation framework, the unnormalized FCF is decoupled into the linear superposition of the LoS and NLoS components. Since the visibility factor $v_{p,q} \in \{0, 1\}$ remains invariant across the frequency shift, the expectation over the visibility state simplifies to its marginal probability. Therefore, the FCF for the LoS component is derived as
\begin{equation}
    R_{\mathrm{LoS}}^F(\Delta f) = \mathbb{E}[v_{p,q}] \frac{K_{p,q}}{K_{p,q}+1} P_{p,q}^{\mathrm{LoS}} \exp\left(j\frac{2\pi \Delta f}{c_0} d_{p,q}^{\mathrm{LoS}}\right).
\end{equation}
Similarly, the FCF for the NLoS components simplifies to
\begin{equation}
\begin{split}
    R_{x}^F(\Delta f) &= \mathbb{E}[v_{p,q}] \frac{1}{K_{p,q}+1} \sum_{n=1}^{N_{x}} \sum_{l=1}^{L_n} P_{p,q,n,l}^{x} \\
    &\quad \times \exp\left(j\frac{2\pi \Delta f}{c_0} D_{p,q,n,l}^{x}\right)
\end{split},
\end{equation}
where the exponential term isolates the phase variations induced by the frequency shift $\Delta f$ over the propagation distance $D_{p,q,n,l}^{x}$.
The derived FCF expressions reveal that the frequency correlation is dictated by the propagation distances of the multipath rays. Since the spherical distances differ across antenna elements, the proposed model shows that the frequency selectivity and the corresponding coherence bandwidth vary significantly across the large-scale array, which highlights the space-frequency non-stationarity of THz XL-MIMO channels.
\subsection{Channel Capacity}
Channel capacity represents the maximum achievable transmission rate of a channel, enabling an effective evaluation of the overall performance of communication systems. Given the frequency selective fading in wideband terahertz communications, the entire bandwidth is uniformly divided into $N_F$ subbands. Assuming that the receiver perfectly estimates the channel matrix while it remains unknown at the transmitter, the channel capacity for each subband can be written as
\begin{equation}\label{eq_capacity}
    C(f_i) = \mathbb{E} \left[ \log_2 \det \left( \mathbf{I} + \frac{\rho_{SNR}}{N_T} \tilde{\mathbf{H}}(f_i) \tilde{\mathbf{H}}^H(f_i) \right) \right],
\end{equation}
where $\mathbf{I}$ is the $N_T \times N_T$ identity matrix, $\rho_{SNR}$ represents the average signal-to-noise ratio (SNR), and $(\cdot)^H$ denotes the conjugate transpose operation.
To ensure a fair evaluation across different channel realizations, the normalized channel matrix $\tilde{\mathbf{H}}(f_i)$ is derived as
\begin{equation}
    \tilde{\mathbf{H}}(f_i) = \frac{\mathbf{H}(f_i)}{\sqrt{\frac{1}{N_T N_R N_F} \sum_{k=1}^{N_F} ||\mathbf{H}(f_k)||_F^2}},
\end{equation}
where $||\cdot||_F$ is the Frobenius norm, and $\mathbf{H}(f_i)\in \mathbb{C}^{N_{T} \times N_{R}}$ is the original channel transfer matrix at the $i$-th frequency point. Here, energy normalization is performed across the entire frequency band rather than at a single frequency point.
\section{Channel Measurement}
\subsection{Channel Measurement System}
\begin{figure}[!t]
\centering
\includegraphics[trim=1mm 1mm 1mm 1mm, clip, width=0.8\columnwidth]{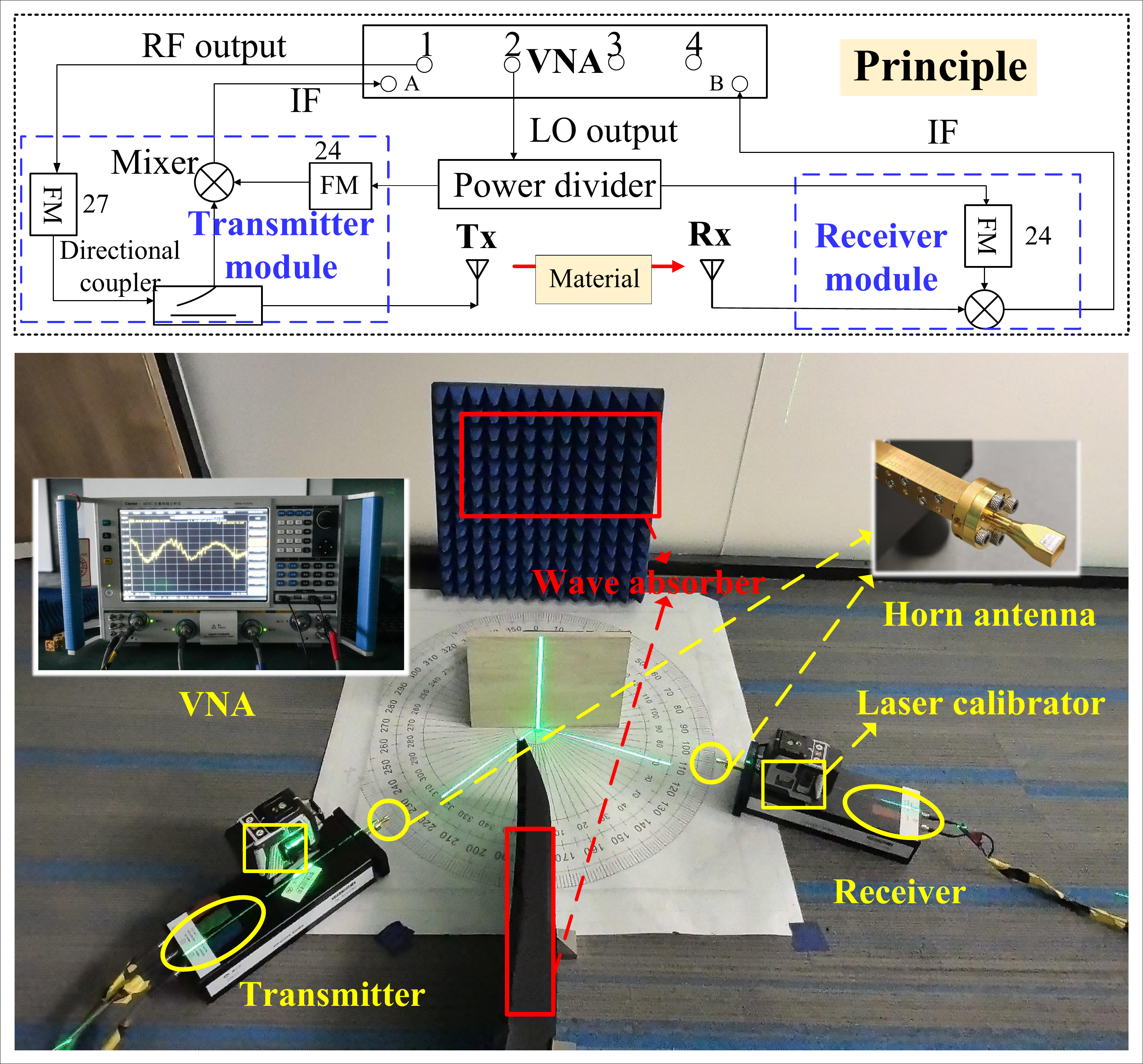}
\caption{Measurement principle and setup of the scattering characteristics.}
\vspace{-2mm}
\label{sca_env}
\end{figure}

To empirically validate the proposed scattering-based channel model, a VNA-based frequency-domain channel measurement system is established \cite{zhang2025indoor}, as shown in Fig. \ref{sca_env}. In this setup, the VNA generates a radio frequency (RF) signal ranging from 9.62 GHz to 14.81 GHz. This signal is delivered to the transmitter module, where it is up-converted to the 260–400 GHz band by an embedded $\times 27$ frequency multiplier. Then, the up-converted signal is mixed with a local oscillator (LO) signal, which sweeps between 10.83 GHz and 16.66 GHz and is multiplied by a factor of 24, to produce a 279 MHz reference intermediate frequency (IF) signal. At the receiver, the propagated signal is acquired and down-converted to produce the corresponding IF test signal. By comparing the test and reference signals, the VNA extracts the $S_{21}$ parameters, providing both amplitude and phase information. Prior to data processing, a direct-link calibration is performed to eliminate the system frequency responses introduced by the VNA, cables, and antennas from the measured $S_{21}$ parameters. Then, the resulting calibrated channel transfer functions (CTFs) are transformed into CIRs via the inverse fast Fourier transform (IFFT).

\begin{figure}[!t]
\centering
\includegraphics[trim=1mm 1mm 1mm 1mm, clip, width=0.8\columnwidth]{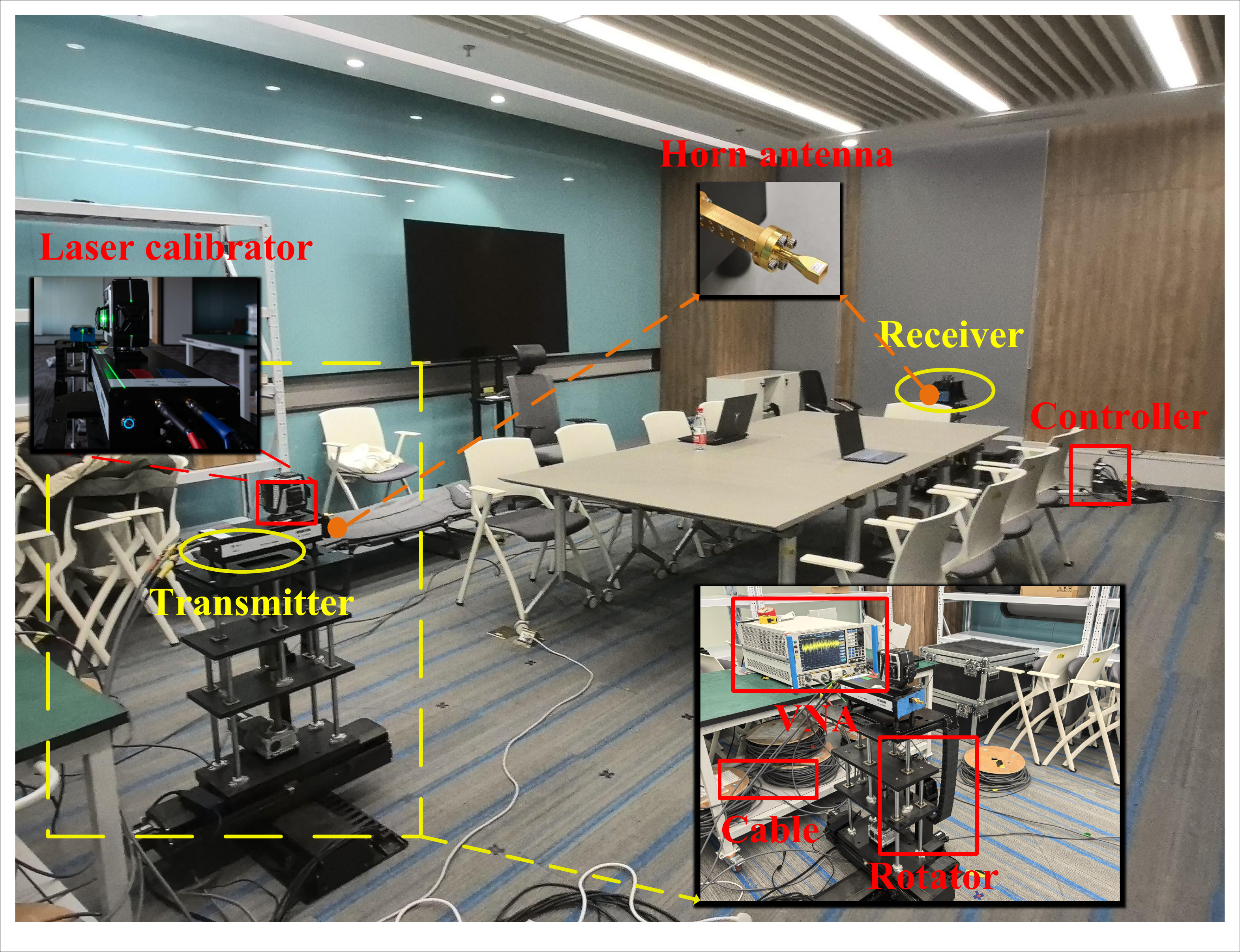}
\\[5pt]
    
\includegraphics[trim=1mm 2mm 10mm 1mm, clip, width=0.8\columnwidth]{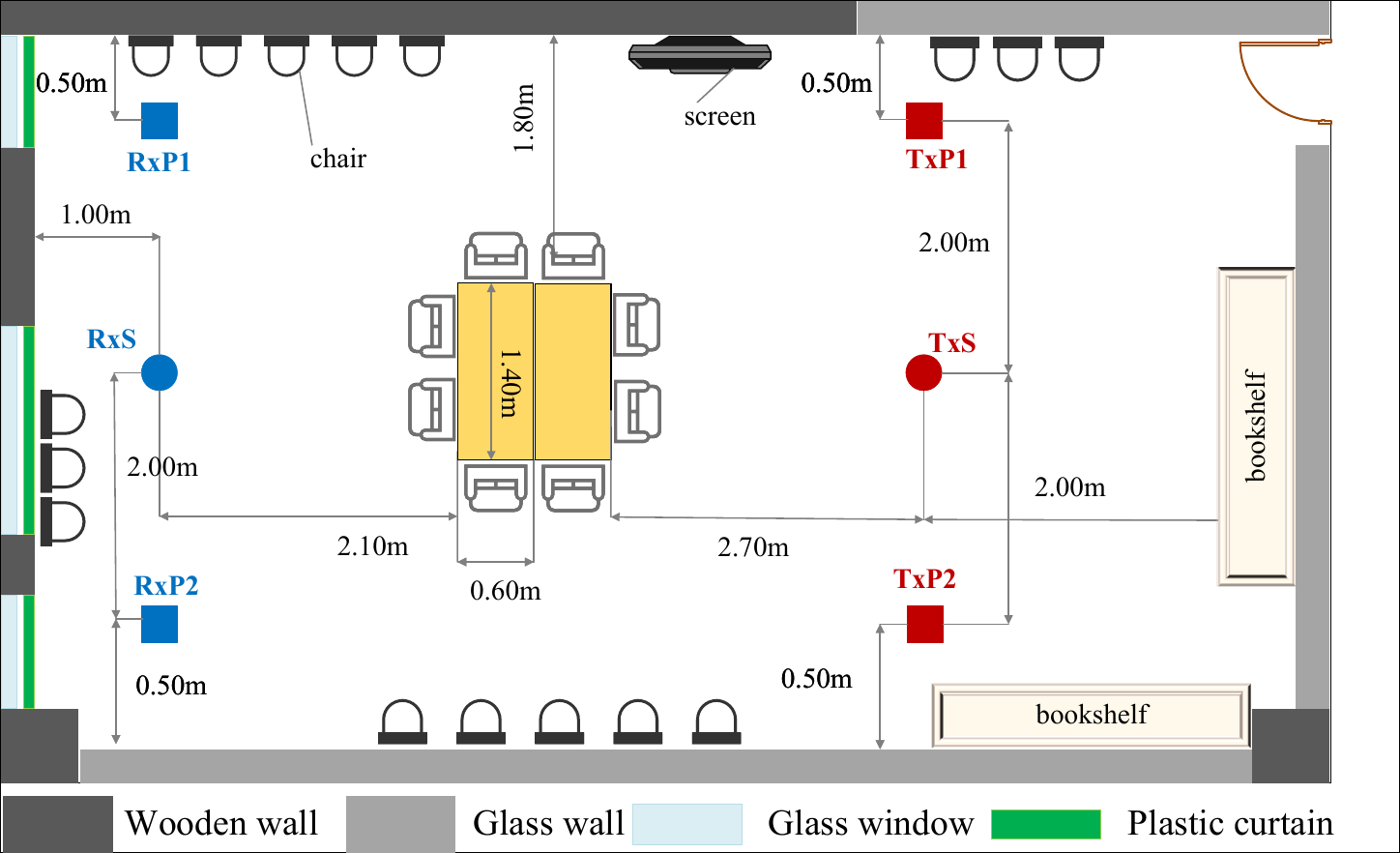}
\caption{Measurement layout and photograph of the indoor meeting room environment.}
\vspace{-2mm}
\label{indoor_env}
\end{figure}

\subsection{Measurement Scenario and Setup}
Three measurement campaigns were conducted across three distinct cases. Case I focuses on measuring the scattering coefficients of three typical materials to validate the scattering theory and derive the improved B-K scattering model. By synthesizing a massive 630-element virtual ULA at the Tx, Case II investigates the spatial non-stationarity of THz channels within the 330–360 GHz band in a typical indoor meeting room. Finally, Case III performs a $128 \times 4$ MIMO measurement campaign in the same environment to analyze the statistical properties and system performance of the proposed model. The detailed setups for each case are provided as follows.
\begin{figure*}[htbp]
    \centering
    \includegraphics[width=0.8\textwidth]{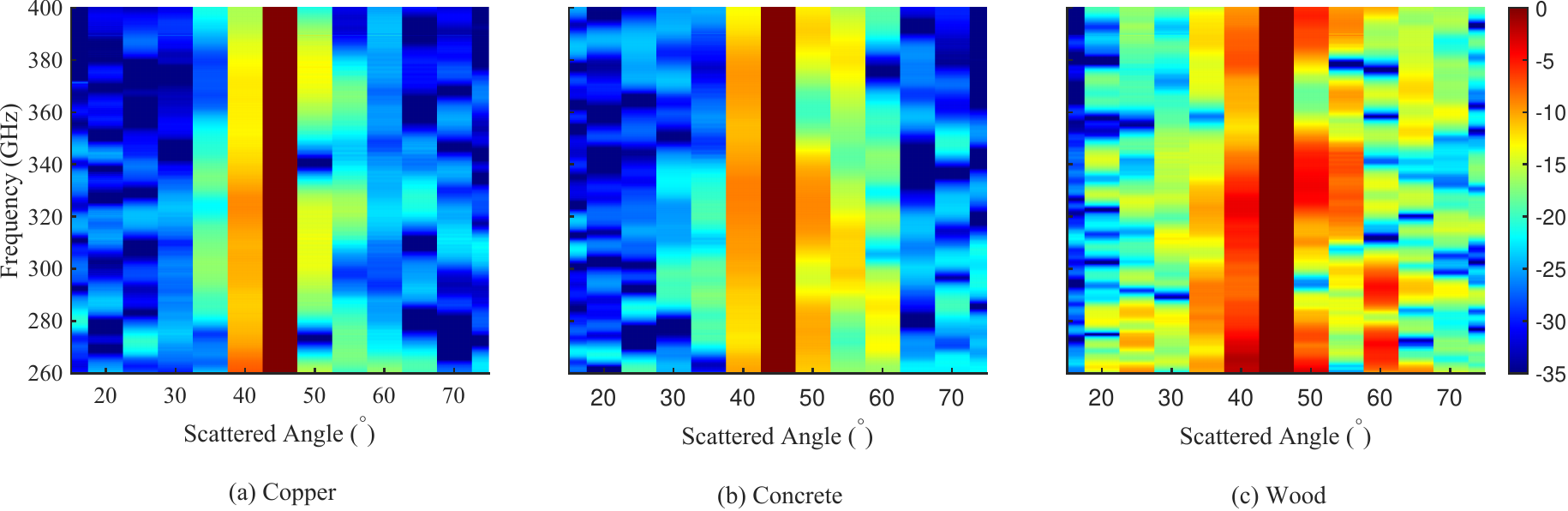}
    \caption{Measured normalized scattered power distributions across the 260–400 GHz band at different scattered angles for (a) copper, (b) concrete, and (c) wood surfaces.}
    \vspace{-2mm}
    \label{Heatmap}
\end{figure*}
\subsubsection{Case 1}
Fig. \ref{sca_env} illustrates the measurement environment and setup for Case I. The measurements are performed on three distinct materials: copper, wood board, and concrete. The material under test is positioned at the center, with the Tx and Rx deployed at the incident and scattering sides, respectively. To satisfy the far-field condition, the Tx and Rx antennas are directed toward the center of the material at a constant radius of 40 cm. Furthermore, radar absorbing material is placed to suppress spurious multipath interference. To ensure angular precision, laser alignment tools are employed at both the Tx and Rx. The measurements are conducted in the transverse electric (TE) mode across the 260-400 GHz frequency range, during which the incident angle is fixed at 45°, while the receiving angle ranges from 15° to 75° in steps of 5°. 
\subsubsection{Case 2}
Case 2 is conducted in a typical indoor meeting room, with its measurement layout and photograph shown in Fig. \ref{indoor_env}.
The dimensions of the room are 10 m $\times$ 5.5 m $\times$ 3.2 m. The Tx and Rx, denoted by red and blue spheres in the figure, are positioned on opposite sides of the room, with their beams directly aligned. Both the Tx and Rx are mounted at an identical height of 0.9 m. Operating within the 330–360 GHz band, a massive 630-element virtual ULA is synthesized at the Tx, while a single antenna is employed at the Rx. To ensure the validity of a quasi-static environment assumption throughout the measurement, all scatterers remain strictly stationary, with no interference from human movement or other dynamic scatterers.
\subsubsection{Case 3}
Case III is conducted in the same indoor meeting room, which is also illustrated in Fig. \ref{indoor_env}. The Tx and Rx, denoted by red and blue squares, are positioned at the corners of the room, with their beams directly aligned. As in Case II, the operating frequency and antenna heights remain unchanged. In this configuration, a 128-element virtual ULA is synthesized at the Tx, while a 4-element virtual ULA is employed at the Rx. To comprehensively extract multipath components (MPCs), the Rx performs a systematic azimuthal rotation from -10° to 10° during the measurement. 

The specific configurations and parameters for each measurement case are summarized in Table~\ref{tab:model_parameters}.
\begin{table}[htbp]
    \centering
    \caption{Measurement Configurations and Parameters}
    \label{tab:model_parameters}
    
    \renewcommand{\arraystretch}{1.3}
        \setlength{\tabcolsep}{3.5pt}

    \begin{tabular}{c|c|c|c}
        \hline
        Parameters & \multicolumn{3}{c}{Values} \\ \hline
        
        Cases & 1 & 2 & 3 \\ \hline
        
        Frequency band & 260--400 GHz & 330--360 GHz & 330--360 GHz \\ \hline
        
        Tx array & Single antenna & 630$\times$1 & 128$\times$1 \\ \hline

        Rx array & Single antenna & Single antenna & 4$\times$1 \\ \hline
        
        Antenna gain at Tx/Rx & 25 dBi & 25 dBi & 25 dBi \\ \hline
        
        HPBW & 10$^\circ$ & 10$^\circ$ & 10$^\circ$ \\ \hline
        
        Sweeping points & 12001 & 3001 & 3001 \\ \hline
        Average noise floor & -145 dBm & -145 dBm & -145 dBm \\ \hline
        Test power & 0.5 mW & 0.5 mW & 0.5 mW \\ \hline
    \end{tabular}
    \vspace{-2mm}
\end{table}
\section{Measurement Results}
\subsection{Scattering Characteristics}
In the terahertz bands, indoor building materials traditionally considered smooth at lower frequencies become electromagnetically rough, as their surface height variations are comparable to the incident wavelengths. When an electromagnetic wave is incident on such a rough surface, it generally scatters in all directions. Specifically, the scattered field can be decomposed into a coherent component reflecting along the specular direction and an incoherent component spreading into other non-specular directions. As the surface roughness increases, the coherent specular reflection becomes heavily attenuated, and the incoherent diffuse scattering dominates the overall reflection behavior. To characterize the reflection behavior, we conduct a comprehensive analysis on the scattering characteristics of three typical indoor materials.
\begin{figure}[!t]
\centering
\includegraphics[width=0.8\columnwidth]{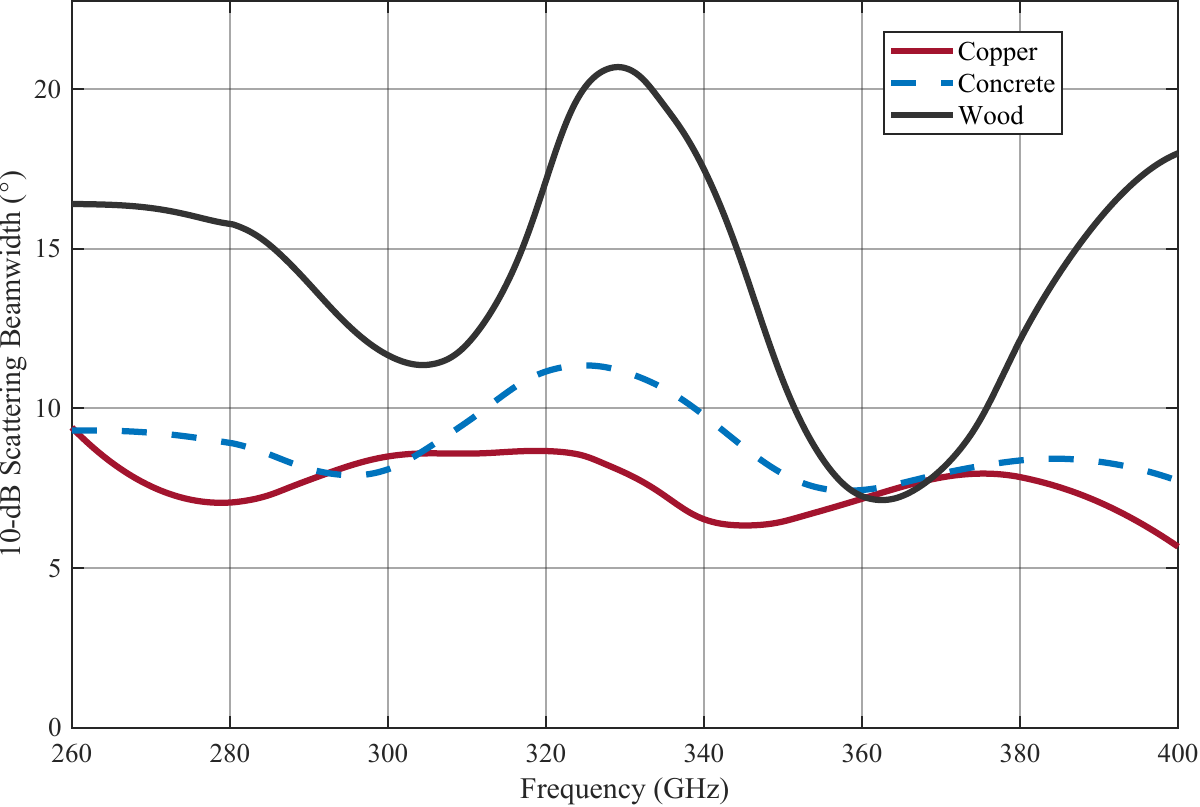}
\caption{The 10-dB scattering beamwidth $\Delta\Theta_{\mathrm{10dB}}$ of copper, concrete, and wood at 260–400 GHz.}
\vspace{-2mm}
\label{ANG}
\end{figure}
\begin{figure*}[htbp]
    \centering
    \includegraphics[width=0.8\textwidth]{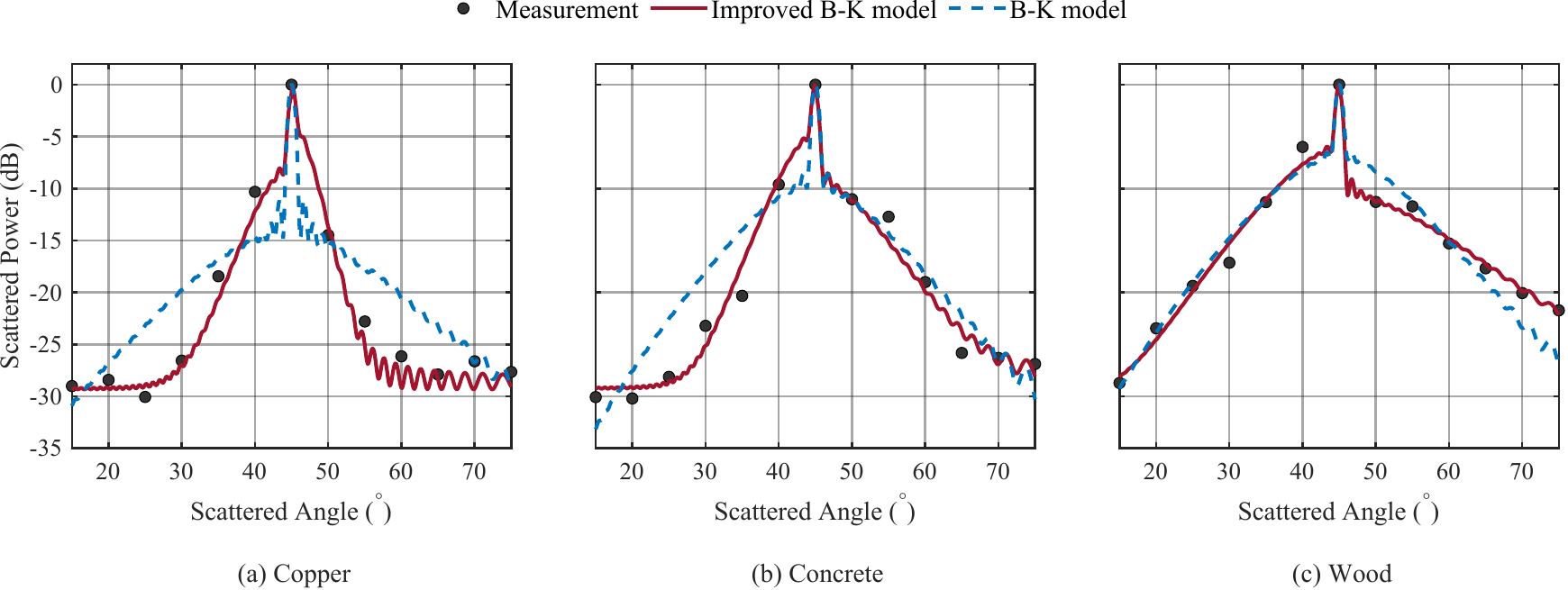}
    \caption{Comparison of measured scattered power and different theoretical models at 345 GHz for (a) copper, (b) concrete, and (c) wood.}
    \label{PAP}
\end{figure*}
Fig. \ref{Heatmap} presents the measured scattering power heatmap across various scattering angles and frequencies, which is normalized relative to the peak power at each frequency bin to remove the frequency-dependent effects of path loss and reflection coefficients. It can be observed that the scattered power gradually attenuates as the scattering angle deviates from the specular direction on both sides. And materials with higher surface roughness exhibit a slower power attenuation rate, leading to a considerably broadened scattering profile. This angular broadening occurs because the increased surface roughness generates a substantial incoherent component, resulting in a strong diffuse scattering effect that spreads the energy across a wider range of angles rather than purely along the specular reflection path. Therefore, the 10-dB scattering beamwidth, denoted as $\Delta\Theta_{\mathrm{10dB}}$, is introduced to characterize the spatial angular dispersion of the scattered power for different materials. Let $\theta_{\mathrm{spec}}$ represent the specular reflection angle and $P_{\mathrm{max}}$ be the peak scattered power. The $\Delta\Theta_{\mathrm{10dB}}$ is formulated as the angular width between the left and right boundaries, $\theta_L$ and $\theta_R$, where the scattered power drops by 10 dB relative to $P_{\mathrm{max}}$. 

Fig. \ref{ANG} shows the measured $\Delta\Theta_{\mathrm{10dB}}$ results for the three materials from 260 to 400 GHz. As observed, the copper exhibits a small and stable $\Delta\Theta_{\mathrm{10dB}}$, fluctuating slightly between $6^\circ$ and $9^\circ$, which confirms that copper acts as a highly specular reflector. In contrast, concrete has the moderate scattering beamwidth ranging from approximately $7^\circ$ to $11^\circ$. The wood shows the significantly broadened $\Delta\Theta_{\mathrm{10dB}}$ that heavily fluctuates with frequency, reaching a peak of over $20^\circ$ near 330 GHz. It indicates that surface roughness not only increases the spatial angular dispersion of power but also introduces high frequency selectivity. Consequently, it is necessary to divide the entire bandwidth into narrower subbands for wideband THz channels.

\begin{figure}[!t]
\centering
\includegraphics[width=0.7\columnwidth]{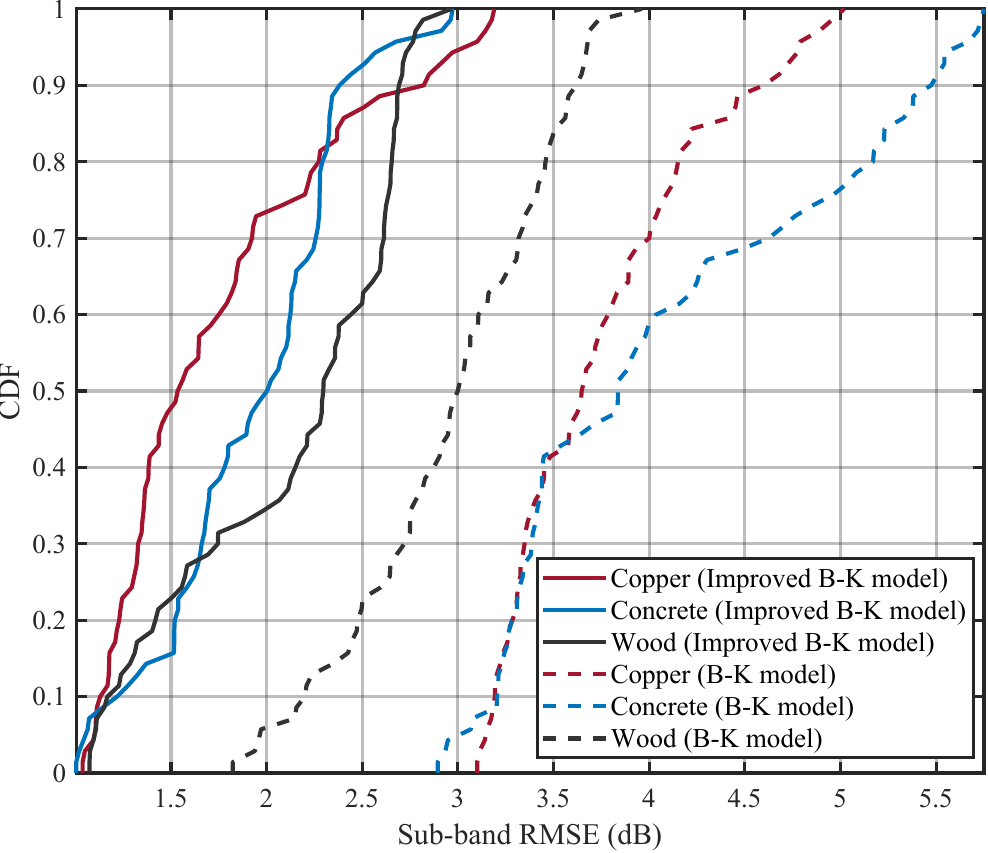}
\caption{CDFs of the sub-band RMSE for the traditional and improved B-K models across different materials in the 260–400 GHz band.}
\vspace{-2mm}
\label{CDF_RMSE}
\end{figure}

Another key observation drawn from Fig. \ref{Heatmap} is that the decay rate of the scattered power is different on either side of the specular reflection direction. And the scattered power gradually stabilizes at larger scattering angles for materials with low surface roughness, which is attributed to the constant diffuse scattering floor introduced by environmental scattering. Accordingly, we propose an improved B-K scattering model that adopts different surface correlation lengths for the two sides of the specular reflection direction, while adding a diffuse background power factor into the power scattering coefficient formulation. Fig. \ref{PAP} illustrates the measured scattered power of different materials at 345 GHz and the fitting results of the two models. It can be seen from Fig. \ref{PAP} that the proposed model provides a much better fit to the measurement data than the traditional model, effectively simulating the differences on either side of the specular reflection as well as the power stabilization at large scattering angles. Furthermore, Fig. \ref{CDF_RMSE} presents the cumulative distribution function (CDF) of the overall RMSE for the measured materials. It can be observed that the overall RMSE of the proposed model is reduced by approximately 2 dB on average when compared with the typical model.

\subsection{Near-field Effect and Spatial Non-Stationarity}
Owing to the extremely short wavelength of the THz band, THz massive MIMO systems can deploy hundreds or even thousands of antenna elements, resulting in a huge spatial aperture. This massive array aperture shifts the operating region, frequently placing the transceivers within the radiative near-field. Consequently, the conventional plane wave and stationarity assumptions at lower frequencies become invalid across the entire array. The channel statistical properties and multipath parameters vary across the antenna elements, exhibiting pronounced near-field effects and spatial non-stationarity.

\begin{figure}[!t]
\centering
\includegraphics[width=0.7\columnwidth]{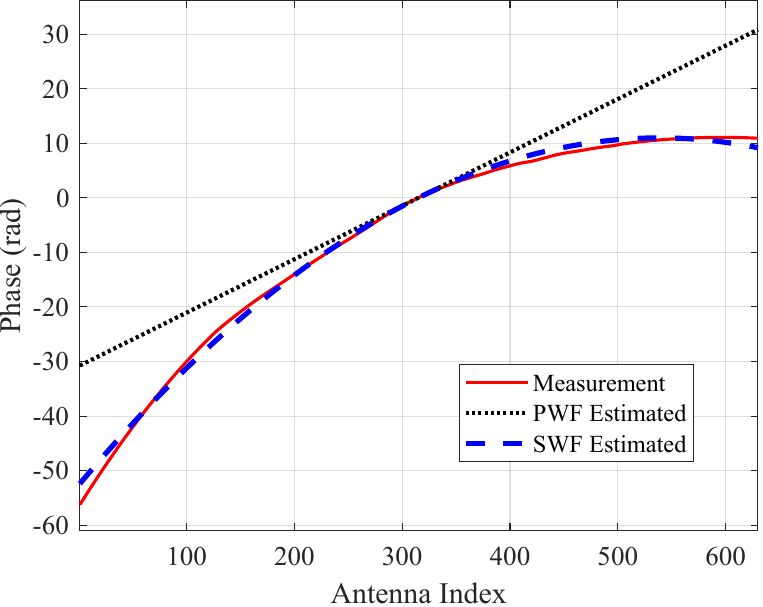}
\caption{Comparison of the measured phase across the antenna array with the estimated phases using the PWF and SWF models.}
\vspace{-2mm}
\label{phase}
\end{figure}

\begin{figure}[!t]
\centering
\includegraphics[width=\columnwidth]{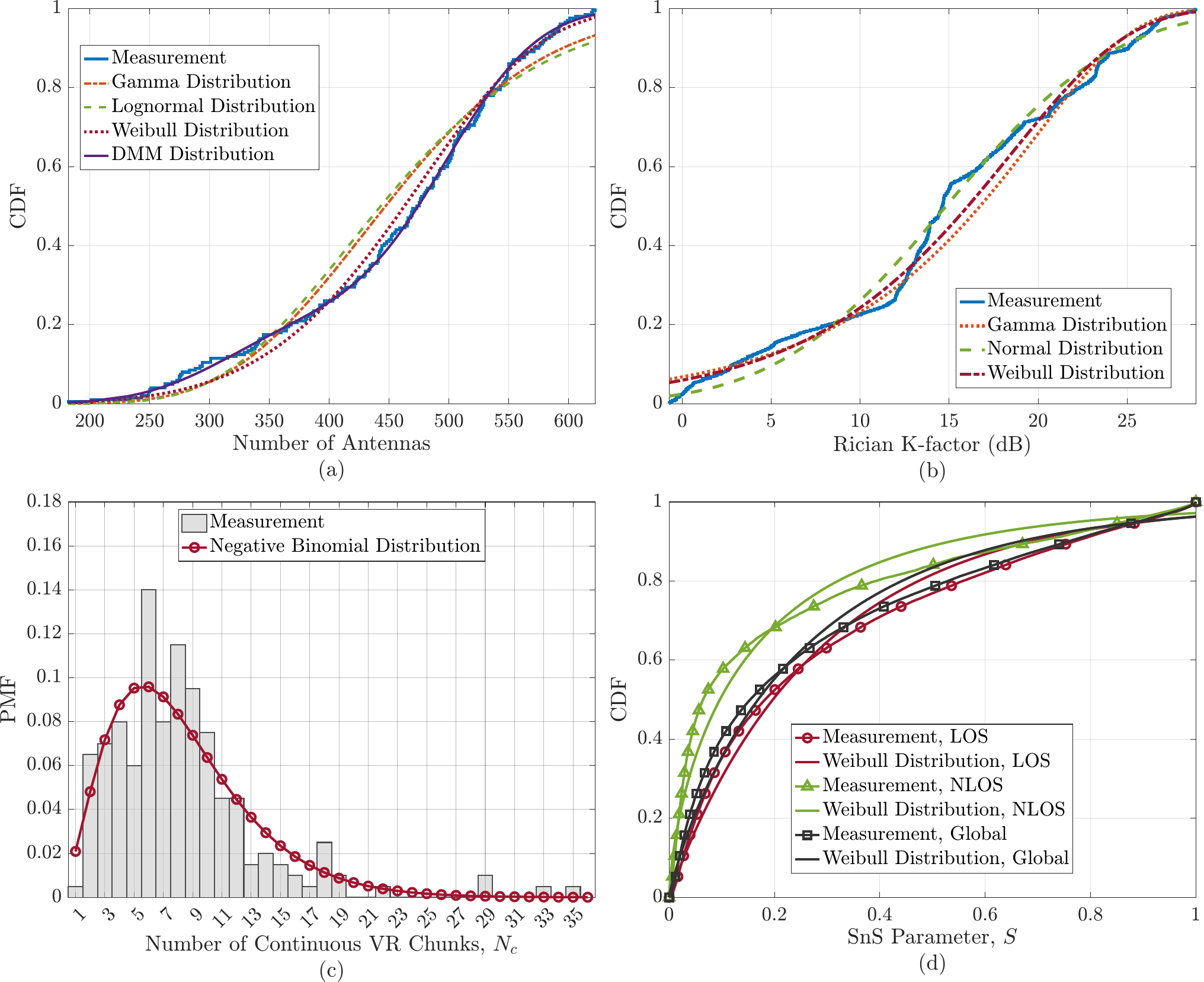}
\caption{Comparison of the empirical distributions of spatial non-stationarity parameters with theoretical models. (a) CDF of the total visibility region lengths. (b) CDF of the Rician K-factor. (c) PMF of the number of continuous VR. (d) CDF of the SnS parameter $S$.}
\vspace{-3mm}
\label{CDF_com}
\end{figure}

Fig. \ref{phase} compares the measured phase across the antenna array with the estimations derived from the plane wavefront (PWF) and spherical wavefront (SWF) models. As observed, the PWF model's linear phase assumption differs significantly from actual measurements at the array edges. Instead, the massive aperture induces a pronounced non-linear phase variation. By adopting the exact Green's function, the SWF model accurately describes this non-linear profile, confirming the necessity of near-field modeling for THz XL-MIMO channels.
\begin{figure}[!t]
\centering
\includegraphics[width=0.8\columnwidth]{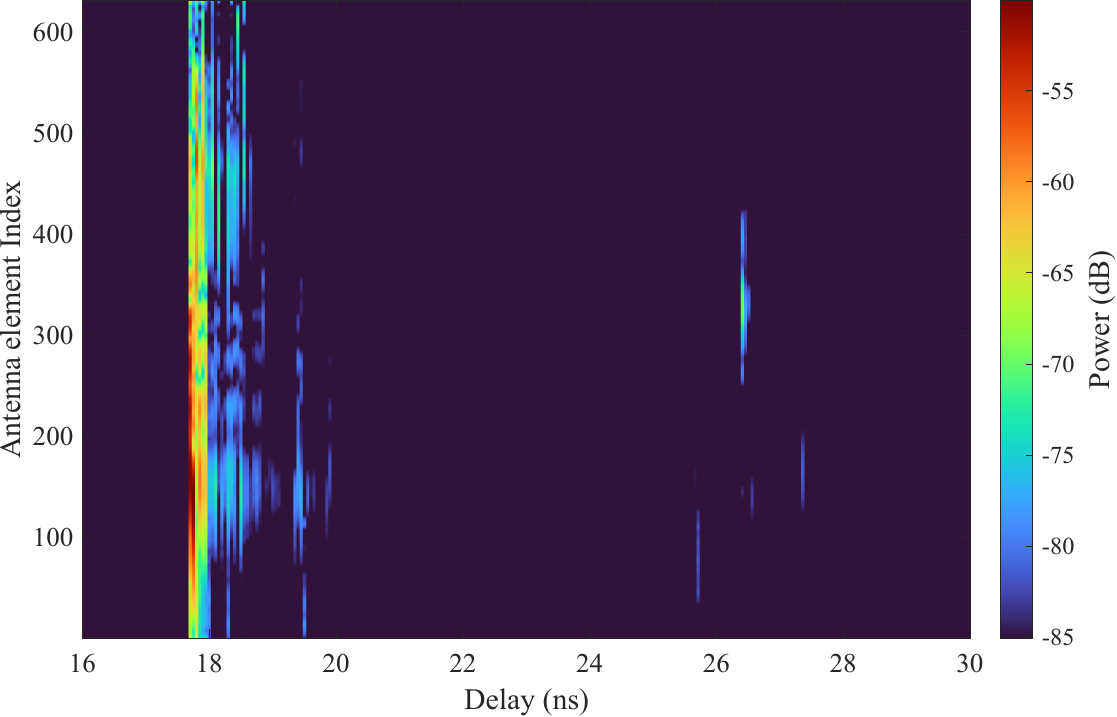}
\caption{PDPs across Tx antenna elements for the TxS-RxS link.}
\vspace{-2mm}
\label{PAD}
\end{figure}
To illustrate the spatial evolution of the channel, a total of 200 MPCs are extracted from the measurement data.  Fig. \ref{PAD} presents the spatial power distribution heatmap, detailing the power distribution of the MPCs across the antenna array. As observed from the heatmap, the SnS of the MPCs exhibits a distinct clustering phenomenon, which is attributed to the THz signals primarily interacting with a few dominant scatterers, whose limited VR lengths make the reflections visible to only certain antenna elements. Fundamentally, the statistical distribution of a visibility region depends on the specific scatterers experienced by the signals. Thus, we adopt the DMM, as previously formulated in Section~\ref{sec:spatial_matrix_formulation}, to fit the statistical distribution of the VR lengths. Fig. \ref{CDF_com} (a) compares the CDF of the VR lengths against the theoretical DMM fitting curve and classical baseline models, including the Exponential, Log-normal, and Weibull distributions. The fitted parameters of the proposed DMM are evaluated as $\omega_1 = \text{0.76}$, $\mu_1 = \text{498.51}$, and $\sigma_1 = \text{60.53}$ for the first component, and $\omega_2 = \text{0.24}$, $\mu_2 = \text{317.96}$, and $\sigma_2 = \text{61.23}$ for the second component. The fitting results show that the DMM has better performance compared to the traditional models. This is because the DMM broadly divides the VRs into long and short length regions, making it better suited to different scattering environments than the traditional single-distribution models.

Fig. \ref{CDF_com} (b) shows the CDF of the Rician K-factors calculated along the antenna array and the corresponding Normal distribution fitting results. The mean and standard deviation of the fitted Normal distribution are $\mu = \text{14.82}$ and $\sigma = \text{7.54}$, respectively.
It can be clearly seen that the Rician K-factors vary from roughly 0 to 25 dB, indicating the significant impact of spatial non-stationarity on the Rician K-factor. Thus, rather than assuming a constant value, independent Rician K-factors should be adopted for individual antenna elements.

While the VR length calculated above is the total VR length of the MPC, the irregular distribution of scatterers in the environment often causes multiple spatial blockages. Therefore, we evaluate the number of continuous VR chunks per path to characterize this discrete behavior. Fig. \ref{CDF_com} (c) shows the probability mass function (PMF) of the number of VR chunks. The distributions can be well approximated by Negative Binomial distributions with the dispersion parameter $r = \text{3.37}$ and the probability parameter $p = \text{0.32}$.

Furthermore, while the multipath birth-death behavior is described by the VR, spatial power non-stationarity persists within the single VR. Thus, we apply maximum normalization to the power across all VRs for each MPC. The distributions of the SnS parameter $S$ are analyzed separately for the overall, LoS, and NLoS paths, as illustrated in Fig. \ref{CDF_com} (d). It is evident that the distributions for all three cases are well approximated by the Weibull distribution. Consequently, to simplify the model, a Weibull distribution with a scale parameter of $A = \text{0.26}$ and a shape parameter of $B = \text{0.88}$ is employed for all MPCs.
\begin{figure}[!t]
    \centering
    \includegraphics[width=\columnwidth]{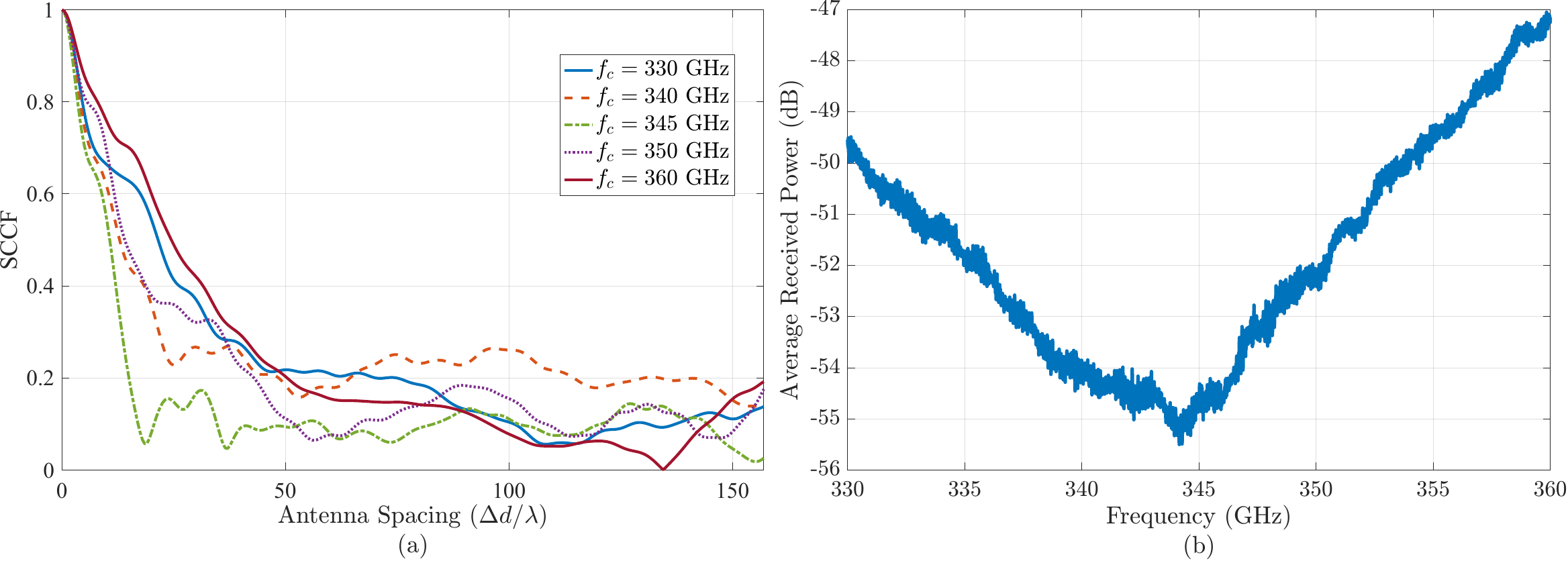}
    \caption{Channel characteristics for the TxP1--RxP1 link within the 330--360~GHz band. (a) measured SCCFs against the antenna spacing ($\Delta d/\lambda$) at different frequencies. (b) averaged received power as a function of frequency.}
    \vspace{-2mm}
    \label{Combined_Fig_SCCF}
\end{figure}

\subsection{Performance Analysis}
To illustrate the spatial and frequency non-stationarity and evaluate the system performance in THz XL-MIMO systems, this subsection analyzes the statistical properties of the proposed channel model, including the SCCF, the FCF, and the channel capacity. Moreover, the analytical results are comprehensively compared with the empirical measurement data. 

\begin{figure*}[!htbp]
    \centering
    \includegraphics[width=\textwidth]{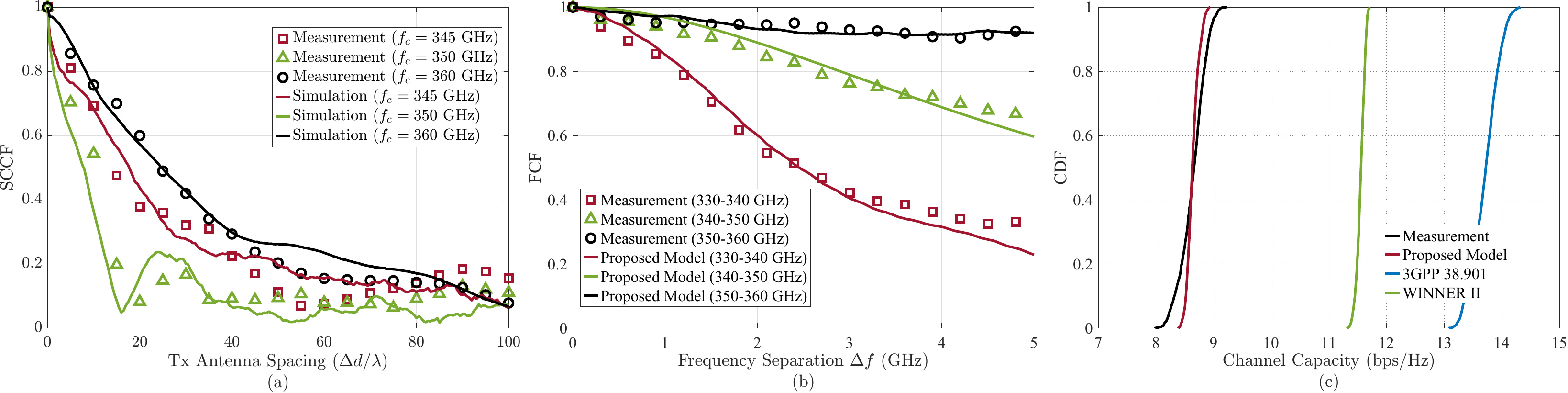}
    \caption{Performance analysis of the proposed channel model for the TxP1-RxP1 link. (a) SCCFs at different frequencies. (b) FCFs across different frequency bands. (c) CDFs of measured and simulated channel capacities for SNR = 15dB.}
    \vspace{-2mm}
    \label{Combined_Fig_sta}
\end{figure*}

Fig. \ref{Combined_Fig_SCCF} (a) illustrates the measured SCCF across five representative frequencies in the indoor office scenario. It is observed that the spatial correlation decays significantly faster at the center frequency than at the band edges, revealing a severe frequency-dependent dispersion. Then the corresponding average received power profile across the entire frequency band is plotted in Fig. \ref{Combined_Fig_SCCF} (b). The power spectrum exhibits a distinct U-shaped trend where the received power peaks at the band edges and undergoes a pronounced attenuation at the center frequencies. It is fundamentally observed that the frequency-dependent fluctuation in the received power profile directly reflects the variation in the VR across the array. At the band edges, the elevated received power corresponds to dense and continuous VRs, indicating that a large subarray shares the same multipath clusters. Consequently, the channel exhibits strong local spatial stationarity, leading to a slow decay in the measured SCCF. Conversely, at the center frequencies where the power undergoes the attenuation, the VR becomes highly sparse which accelerates the spatial decorrelation, causing the rapid drop observed in the SCCF. 

Subsequently, the channel model is configured with the extracted VR parameters, including the number of continuous VR chunks and the visibility lengths at each frequency, and the corresponding SCCF simulation results are depicted in Fig. \ref{Combined_Fig_sta} (a). As observed, the measured data and the simulation results are in good agreement, demonstrating the accuracy and validity of the proposed  channel model.
Fig. \ref{Combined_Fig_sta} (b) illustrates the absolute values of the FCFs for the proposed THz channel model compared with measurement data at different center frequencies, specifically $f_0 = 335$ GHz, $f_1 = 345$ GHz, and $f_2 = 355$ GHz. It can be found that the FCFs exhibit distinct variations across different frequencies, highlighting the frequency-domain non-stationarity in the THz band. Moreover, higher center frequencies exhibit greater FCF values because the increased path loss reduces the number of propagation paths, leading to a sparser scattering environment that produces higher frequency-domain correlation. In addition, the simulation results agree with the measurement data well, which verifies the accuracy of the proposed channel model.

Fig. \ref{Combined_Fig_sta} (c) compares the measured MIMO channel capacity with simulations from various models. As expected, the proposed model simulation results match the measurement data well, while the 3GPP 38.901 and WINNER II models significantly overestimate the channel capacity. This is because these standard models are designed for rich-scattering environments typical of sub-6 GHz and mmWave bands. Thus, they tend to overestimate the number of effective propagation clusters in THz scenarios. By overlooking this spatial sparsity, these models overestimate the spatial degrees of freedom, resulting in increased channel capacity estimations.

\section{Conclusion}
In this paper, a hybrid near-field indoor channel model based on surface scattering characteristics was proposed for THz XL-MIMO systems. The investigation of THz XL-MIMO indoor propagation established that conventional ideal smooth surface assumptions were no longer applicable. Measurements across the 260–400 GHz band revealed that surface roughness of typical building materials introduced severe spatial angular dispersion and frequency selectivity. Consequently, an improved B-K scattering model was developed to characterize the asymmetrical power attenuation around the specular reflection direction and the diffuse background power.
To address the complex propagation mechanisms, the proposed channel model analyzed SB and MB clusters independently. Specifically, it applied deterministic rough surface scattering theory to SB clusters, while using a geometry-statistical method for MB clusters. This structural separation effectively linked deterministic rough surface scattering with statistical multipath geometries.
Furthermore, the near-field spatial non-stationarity measurements from the massive 630-element virtual array in the 330–360 GHz band demonstrated that the evolution of spatial VRs could be characterized by adopting the DMM and the NB distribution. Additionally, these measurement data confirmed that the power fluctuations within each VR were effectively modeled by a Weibull distribution.
To comprehensively evaluate the proposed model, key statistical properties, including the SCCF, FCF, and channel capacity, were theoretically derived and empirically validated against the measured data within the same band. The results demonstrated that the proposed model aligned closely with these observations. By accurately describing the spatial sparsity, the model successfully limited the severe overestimation of channel capacity that typically affected traditional standard models, such as 3GPP 38.901 and WINNER II.

\bibliographystyle{IEEEtran}

\bibliography{references}

@article{wang2024far,
  author={Wang, Yiqin and Sun, Shu and Han, Chong},
  journal={IEEE Communications Letters}, 
  title={Far- and Near-Field Channel Measurements and Characterization in the Terahertz Band Using a Virtual Antenna Array}, 
  month={May.},
  year={2024},
  volume={28},
  number={5},
  pages={1186-1190},
}

@article{aljzari2025subthz,
  author={Al-jzari, Amar and He, Yubei and Hu, Jiahao and Salous, Sana},
  journal={IEEE Transactions on Terahertz Science and Technology}, 
  title={{Sub-THz} and {THz} Channel Measurements and Characteristic Analysis in Indoor and Outdoor Environments for {6G} Wireless Systems}, 
  month={Nov.},
  year={2025},
  volume={15},
  number={6},
  pages={976-984},
}

@article{wang2024propagation,
  author={Wang, Yang and Li, Chuan and Liao, Xi and Yu, Ziming and Wang, Guangjian and Li, Xianjin and Zhang, Jie},
  journal={IEEE Wireless Communications Letters}, 
  title={Propagation Characteristics and Modeling of Monostatic {RCS} Scattering From Indoor Building Material Above 215-{GHz} Frequency Bands}, 
  month={Mar.},
  year={2024},
  volume={13},
  number={3},
  pages={642-646},
}

@article{azpilicueta2023diffuse,
  author={Azpilicueta, Leyre and Schultze, Alper and Celaya-Echarri, Mikel and Rodr{\'i}guez-Corbo, Fidel A. and Constantinou, Costas and Shubair, Raed M. and Falcone, Francisco and Navarro-C{\'i}a, Miguel},
  journal={IEEE Transactions on Antennas and Propagation}, 
  title={Diffuse-Scattering-Informed Geometric Channel Modeling for THz Wireless Communications Systems}, 
  year={2023},
  month={Oct.},
  volume={71},
  number={10},
  pages={8226-8238},
}

@article{serghiou2022terahertz,
  author={Serghiou, Demos and Khalily, Mohsen and Brown, Tim W. C. and Tafazolli, Rahim},
  journal={IEEE Communications Surveys {\&} Tutorials}, 
  title={Terahertz Channel Propagation Phenomena, Measurement Techniques and Modeling for {6G} Wireless Communication Applications: A Survey, Open Challenges and Future Research Directions},
  month={Fourthquarter},
  year={2022},
  volume={24},
  number={4},
  pages={1957-1996},
}

@article{tarboush2021teramimo,
  author={Tarboush, Simon and Sarieddeen, Hadi and Chen, Hui and Loukil, Mohamed Habib and Jemaa, Hatem and Alouini, Mohamed-Slim and Al-Naffouri, Tareq Y.},
  journal={IEEE Transactions on Vehicular Technology}, 
  title={{TeraMIMO}: A Channel Simulator for Wideband Ultra-Massive {MIMO} Terahertz Communications},
  month={Dec.},
  year={2021},
  volume={70},
  number={12},
  pages={12325-12341},
}

@article{wang2023novelthz,
  author={Wang, Jun and Wang, Cheng-Xiang and Huang, Jie and Feng, Rui and Aggoune, el-Hadi M. and Chen, Yunfei},
  journal={IEEE Transactions on Vehicular Technology}, 
  title={A Novel {THz} Massive {MIMO} Beam Domain Channel Model for {6G} Wireless Communication Systems}, 
  month={Aug.},
  year={2023},
  volume={72},
  number={8},
  pages={9704-9719},
}

@article{wang2021novel,
  author={Wang, Jun and Wang, Cheng-Xiang and Huang, Jie and Wang, Haiming and Gao, Xiqi and You, Xiaohu and Hao, Yang},
  journal={IEEE Transactions on Vehicular Technology}, 
  title={A Novel {3D} Non-Stationary {GBSM} for {6G} {THz} Ultra-Massive {MIMO} Wireless Systems}, 
  month={Dec.},
  year={2021},
  volume={70},
  number={12},
  pages={12312-12324},
}

@article{zhang2024deterministic,
  author={Zhang, Jianhua and Lin, Jiaxin and Tang, Pan and Fan, Wei and Yuan, Zhiqiang and Liu, Ximan and Xu, Huixin and Lyu, Yejian and Tian, Lei and Zhang, Ping},
  journal={IEEE Communications Magazine}, 
  title={Deterministic Ray Tracing: A Promising Approach to {THz} Channel Modeling in {6G} Deployment Scenarios}, 
  month={Feb.},
  year={2024},
  volume={62},
  number={2},
  pages={48-54},
}

@inproceedings{priebe2011nonspecular,
  author={Priebe, Sebastian and Jacob, Martin and Jansen, Christian and K{\"u}rner, Thomas},
  booktitle={Proceedings of the 5th European Conference on Antennas and Propagation (EUCAP)}, 
  title={Non-Specular Scattering Modeling for {THz} Propagation Simulations}, 
  month={Apr.},
  year={2011},
  pages={3824-3828},
}

@article{khamse2023scattering,
  author={Khamse, Amir Mehdirezaei and Dong, Xiaodai and Ferdinand, Nuwan},
  journal={IEEE Open Journal of the Communications Society}, 
  title={The Scattering Channel Model for Terahertz Wireless Communications}, 
  month={Mar.},
  year={2023},
  volume={4},
  pages={846-860},
}

@article{sheikh2020study,
  author={Sheikh, Fawad and Gao, Yuan and Kaiser, Thomas},
  journal={IEEE Transactions on Antennas and Propagation}, 
  title={A Study of Diffuse Scattering in Massive {MIMO} Channels at Terahertz Frequencies}, 
  month={Feb.},
  year={2020},
  volume={68},
  number={2},
  pages={997-1008}
}

@article{han2015multi,
  author={Han, Chong and Bicen, A. Ozgur and Akyildiz, Ian F.},
  journal={IEEE Transactions on Wireless Communications}, 
  title={Multi-Ray Channel Modeling and Wideband Characterization for Wireless Communications in the Terahertz Band}, 
  month={May.},
  year={2015},
  volume={14},
  number={5},
  pages={2402-2412}
}

@article{dong2026performance,
  author={Dong, Haofan and Akan, Ozgur B.},
  title={Performance Limits of Hardware-Constrained {THz} Inter-Satellite {MIMO-ISAC} Systems}, 
  journal={IEEE Transactions on Communications}, 
  month={Apr.},
  year={2026},
  doi={10.1109/TCOMM.2026.3686700},
  note={Early Access}
}

@article{isac2024beam,
  author={Chen, Wenrong and Li, Lingxiang and Chen, Zhi and Liu, Yuanwei and Ning, Boyu and Quek, Tony Q. S.},
  journal={IEEE Transactions on Vehicular Technology}, 
  title={{ISAC}-Enabled Beam Alignment for Terahertz Networks: Scheme Design and Coverage Analysis}, 
  month={Dec.},
  year={2024},
  volume={73},
  number={12},
  pages={19019-19033}
}

@article{lee2026multilevel,
  author={Lee, Tae-Ju and Singh, Pankaj and Jung, Sung-Yoon},
  journal={IEEE Access}, 
  title={Multilevel Bipolar Pulse-Position Modulation With Level Trimming for Graphene-Based Terahertz Links in Wireless Network-on-Chips}, 
  month={Mar.},
  year={2026},
  volume={14},
  pages={45320-45337},
}

@ARTICLE{chen2024novel,
  author={Ali, Amjad and Aazam, Mohammad and Al Dabel, Maryam M. and Ali, Farman and Bashir, Ali Kashif and El-Sappagh, Shaker and Al-Fuqaha, Ala},
  journal={IEEE Consumer Electronics Magazine}, 
  title={Novel Data Fusion Scheme for Enhanced User Experiences in Terahertz-Enabled {IoNT}}, 
  month={Sep.},
  year={2025},
  volume={14},
  number={2},
  pages={90-102},
  }

@INPROCEEDINGS{qiu2024terahertz,
  author={Aqlan, Basem and Kubiczek, Tobias and Balzer, Jan C.},
  booktitle={2025 International Conference on Mobile and Miniaturized Terahertz Systems (ICMMTS)}, 
  title={Terahertz Radar Cross Section Measurement and Digital Twin Simulation for Advanced Insect Monitoring},
  month={Feb.},
  year={2025},
  volume={},
  number={},
  pages={1-3},
  }

@article{wang2023road,
  author={Wang, Cheng-Xiang and You, Xiaohu and Gao, Xiqi and Zhu, Xu and Zong, Zhaocheng and Zhang, Chuan and Wang, Hua and Huang, Yongming and Chen, Yunfei and Haas, Harald},
  journal={IEEE Communications Surveys \& Tutorials}, 
  title={On the Road to 6G: Visions, Requirements, Key Technologies, and Testbeds}, 
  month={Secondquarter},
  year={2023},
  volume={25},
  number={2},
  pages={905-974}
}

@article{lyu2026environment,
  author={Lyu, Yejian and Yuan, Zhiqiang and Wymeersch, Henk and Han, Chong},
  journal={IEEE Transactions on Wireless Communications}, 
  title={Environment-Aware Channel Measurement and Modeling for Terahertz Monostatic Sensing}, 
  month={Mar.},
  year={2026},
  volume={25},
  number={},
  pages={13369-13380}
}

@article{chowdhury20206g,
  author={Chowdhury, Mostafa Zaman and Shahjalal, Md. and Ahmed, Shakil and Jang, Yeong Min},
  journal={IEEE Open Journal of the Communications Society}, 
  title={{6G} Wireless Communication Systems: Applications, Requirements, Technologies, Challenges, and Research Directions}, 
  month={Jul.},
  year={2020},
  volume={1},
  number={},
  pages={957-975}
}

@article{tataria20216g,
  author={Tataria, Harsh and Shafi, Mansoor and Molisch, Andreas F. and Dohler, Mischa and Sj{\"o}land, Henrik and Tufvesson, Fredrik},
  journal={Proceedings of the IEEE}, 
  title={{6G} Wireless Systems: Vision, Requirements, Challenges, Insights, and Opportunities}, 
  month={Jul.},
  year={2021},
  volume={109},
  number={7},
  pages={1166-1199}
}

@article{wang20206g,
  author={Wang, Cheng-Xiang and Huang, Jie and Wang, Haiming and Gao, Xiqi and You, Xiaohu and Hao, Yang},
  journal={IEEE Vehicular Technology Magazine}, 
  title={{6G} Wireless Channel Measurements and Models: Trends and Challenges}, 
  month={Dec.},
  year={2020},
  volume={15},
  number={4},
  pages={150-158}
}

@article{han2022terahertz,
  author={Han, Chong and Wang, Yiqin and Li, Yuanbo and Chen, Yi and Abbasi, Naveed A. and K{\"u}rner, Thomas and Molisch, Andreas F.},
  journal={IEEE Communications Surveys \& Tutorials}, 
  title={Terahertz Wireless Channels: A Holistic Survey on Measurement, Modeling, and Analysis}, 
  month={Thirdquarter},
  year={2022},
  volume={24},
  number={3},
  pages={1670-1707}
}

@article{zhou2026general,
  author={Zhou, Zihao and Wang, Cheng-Xiang and Huang, Jie and Xin, Lijian and Zhang, Li and Aggoune, El-Hadi M.},
  journal={IEEE Transactions on Wireless Communications}, 
  title={A General {6G} Cross-Band Channel Model Toward Standardization Verified by 0.7--39-{GHz} Channel Measurements}, 
  month={},
  year={2026},
  volume={25},
  number={},
  pages={12422-12437}
}

@article{rappaport2019wireless,
  author={Rappaport, Theodore S. and Xing, Yunchou and Kanhere, Ojas and Ju, Shihao and Madanayake, Arjuna and Mandal, Soumyajit and Alkhateeb, Ahmed and Trichopoulos, Georgios C.},
  journal={IEEE Access}, 
  title={Wireless Communications and Applications Above 100 {GHz}: Opportunities and Challenges for {6G} and Beyond}, 
  month={Jun.},
  year={2019},
  volume={7},
  number={},
  pages={78729-78757}
}

@ARTICLE{Shafie2023terahertz,
  author={Shafie, Akram and Yang, Nan and Han, Chong and Jornet, Josep Miquel and Juntti, Markku and Kürner, Thomas},
  journal={IEEE Network}, 
  title={Terahertz Communications for 6G and Beyond Wireless Networks: Challenges, Key Advancements, and Opportunities}, 
  month={May/Jun.},
  year={2023},
  volume={37},
  number={3},
  pages={162-169}
  }

@techreport{latvaaho2019key,
  author={Latva-aho, Matti and Leppanen, Kari},
  institution={6G Flagship, University of Oulu, Finland}, 
  title={Key Drivers and Research Challenges for {6G} Ubiquitous Wireless Intelligence}, 
  month={Sep.},
  year={2019},
  volume={},
  number={},
  pages={1-77}
}

@techreport{itu2023imt2030,
  author={{Recommendation ITU-R M.2160-0}},
  title={Framework and overall objectives of the future development of {IMT} for 2030 and beyond},
  institution={ITU-R},
  year={2023},
  month={Nov.}
}

@article{kim2026power,
  author={Kim, Yongwan and Kim, Hooyoung and Jo, Junpyo},
  journal={IEEE Transactions on Antennas and Propagation}, 
  title={Power-Conserving Propagation Graph Modeling for Diffuse Scattering Analysis via Rigorous Power Integration}, 
  month={Apr.},
  year={2026},
  volume={74},
  number={4},
  pages={3320-3334}
}

@article{you2021towards,
  author={You, Xiaohu and Wang, Cheng-Xiang and Huang, Jie and Gao, Xiqi and Zhang, Zhaoyang and Wang, Minghua and Huang, Yongming and Zhang, Chuan and Jiang, Xiaohuan and Wang, Jinyuan},
  journal={Science China Information Sciences}, 
  title={Towards 6G wireless communication networks: vision, enabling technologies, and new paradigm shifts}, 
  month={Jan.},
  year={2021},
  volume={64},
  number={11},
  pages={110301}
}

@ARTICLE{jiang2024terahertz,
  author={Jiang, Wei and Zhou, Qiuheng and He, Jiguang and Habibi, Mohammad Asif and Melnyk, Sergiy and El-Absi, Mohammed and Han, Bin and Renzo, Marco Di and Schotten, Hans Dieter and Luo, Fa-Long and El-Bawab, Tarek S. and Juntti, Markku and Debbah, Mérouane and Leung, Victor C. M.},
  journal={IEEE Communications Surveys {\&} Tutorials}, 
  title={Terahertz Communications and Sensing for {6G} and Beyond: A Comprehensive Review},
  month={Fourthquarter},
  year={2024},
  volume={26},
  number={4},
  pages={2326-2381}
}

@ARTICLE{xu2025spatial,
  author={Xu, Huixin and Zhang, Jianhua and Tang, Pan and Xing, Hongbo and Han, Chong and Tian, Lei and Wang, Qixing and Liu, Guangyi},
  journal={IEEE Transactions on Wireless Communications}, 
  title={Empirical Study on Near-Field and Spatial Non-Stationarity Modeling for {THz} {XL-MIMO} Channel in Indoor Scenario}, 
  month={Sep.},
  year={2026},
  volume={25},
  number={},
  pages={4435-4451}
  }

@ARTICLE{taleb2023scattering,
  author={Taleb, Fatima and Hernandez-Cardoso, Goretti G. and Castro-Camus, Enrique and Koch, Martin},
  journal={IEEE Transactions on Terahertz Science and Technology}, 
  title={Transmission, Reflection, and Scattering Characterization of Building Materials for Indoor {THz} Communications}, 
  month={Sep.},
  year={2023},
  volume={13},
  number={5},
  pages={421-430}
  }

@book{beckmann1987scattering,
  author={Beckmann, Petr and Spizzichino, Andre},
  title={The Scattering of Electromagnetic Waves from Rough Surfaces},
  publisher={Artech House Radar Library},
  year={1987}
}

@ARTICLE{zhang2025indoor,
  author={Zhang, Taihao and He, Yongchao and Pan, Cunhua and Ren, Hong and Chang, Hengtai and Wang, Cheng-Xiang and Wang, Jiangzhou},
  journal={IEEE Antennas and Wireless Propagation Letters}, 
  title={Indoor Channel Measurements and Characterization for Virtual Multiantenna at 260 {GHz} to 400 {GHz}}, 
  month={Nov.},
  year={2025},
  volume={24},
  number={11},
  pages={3841-3845},
}

@ARTICLE{priebe2013power,
  author={Priebe, Sebastian and K{\"u}rner, Thomas},
  journal={IEEE Transactions on Wireless Communications}, 
  title={Stochastic Modeling of {THz} Indoor Radio Channels},
  month={Sep.},
  year={2013},
  volume={12},
  number={9},
  pages={4445-4455}
  }

@ARTICLE{wang2021angle,
  author={Wang, Jun and Wang, Cheng-Xiang and Huang, Jie and Wang, Haiming and Gao, Xiqi},
  journal={IEEE Journal on Selected Areas in Communications}, 
  title={A General 3D Space-Time-Frequency Non-Stationary {THz} Channel Model for {6G} Ultra-Massive {MIMO} Wireless Communication Systems},
  month={Jun.},
  year={2021},
  volume={39},
  number={6},
  pages={1576-1589},
  }
\end{document}